\documentclass[showkeys, pra,aps,twocolumn,twoside,showpacs]{revtex4-2}
\usepackage[T1]{fontenc}
\usepackage{amsfonts}
\usepackage{amssymb}
\usepackage{mathrsfs}
\usepackage[intlimits]{amsmath}
\usepackage{bm, bbm}
\usepackage[dvips]{graphicx}

\usepackage[T1]{fontenc}
\usepackage[latin2]{inputenc}
\usepackage {hyperref}
\usepackage{amsmath}
\usepackage{latexsym}

\usepackage{color}
\usepackage[svgnames]{xcolor} 
\usepackage{soul}
\usepackage{graphicx}
\usepackage[export]{adjustbox}
\usepackage{subcaption}
\usepackage{float}

\def\be{\begin{equation}}
\def\ee{\end{equation}}
\def\ben{\begin{eqnarray}}
\def\een{\end{eqnarray}}

\hypersetup{
    colorlinks=true,
    linkcolor=blue,
    filecolor=magenta,      
    urlcolor=blue,
    pdftitle={Overleaf Example},
    pdfpagemode=FullScreen,
    }
\begin{document}
\title{Objectivity in the quantum Brownian motion revisited}
\author{Tae-Hun Lee}\email{taehunee@gmail.com}
\affiliation{\href{https://www.cft.edu.pl/}{Center for Theoretical Physics}, Polish Academy of Sciences,\\ Aleja Lotnik\'ow 32/46, 02-668 Warsaw, Poland}
%\author{Jaros\l{}aw K. Korbicz}\email{Contact author: jkorbicz@cft.edu.pl}
\date{\today}

\begin{abstract}
In this article we revisit objectivity conditions for the quantum Brownian motion (QBM) model under the recoilless (Born-Oppenheimer) limit. The purpose of this study is to correct and clarify the previous objectivity analysis based on the spectrum broadcast structure (SBS). We find that the objectivity for QBM with the finite number of the environments cannot be achieved completely but only depend on a timescale.
%Tuziemski and Korbicz (2015, Europhys. Lett., 112, 40008)
We show that a system with the finite number of environmental oscillators can the objectivity only with respect to the associated timescales defined by frequency relation between a central oscillator and environmental oscillators. In addition, our analysis of the influence of a oscillator trajectory on objectivity answers the previous unsolved question why the objectivity is enhanced as the phase gets closer to $\pi/2$.
\end{abstract}


\maketitle
%%%%%%%%%%%%%%%%%%%%%%%%%%%%%%%%%%%%%%%%%%%%%%%%%%%%%%%%%%%%%%%%%%%%%%%%%%%%%%%%%%%%%%%
\section{Introduction}\label{Introduction}
We undoubtedly believe that quantum mechanics is the fundamental law of nature which encompasses classical physics. However, ironically, since quantum mechanics was born more than a century ago it has remained unsolved to completely explain how a general quantum state of a macroscopic object unitarily evolves into a classical state seemingly having a definite observable. Before answering the question whether quantum mechanics needs to be modified due to this difficulty, it would be more realistic to first ask how far quantum mechanics can reach the classicality. 

Quantum Darwinism and its possible quantum state realization, called the spectrum broadcast structures (SBS) \cite{ZurekNature2009,ZurekPhysToday2014,Korbicz:2014,Korbicz:2020tkf} are physical mechanisms explaining the emergent classicality in quantum mechanics, especially objectivity through decoherence due to environmental interactions. Objectivity is expressed by the statement  ``A state of the system $S$ exists objectively if many observers can determine the state of $S$ independently, and without perturbing it.'' \cite{Korbicz:2020tkf}.
Quantum Darwinism finds a process that quantum states achieve such objective classicality through interaction with environment, producing information proliferation into the environment and creating simultaneously unique observables called ``pointer states''. Mathematically, this is expressed as the common information between a system($S$) and the macro fractional environment($fE$), i.e. the quantum mutual information of the system-composite system, $I(\rho_{S:fE})$, reaches the information of a system, i.e. von Neumann entropy of a system, $H(\rho_S)$, i.e. $I(\rho_{S:fE} )=H(\rho_S)$. Objectivity based on quantum Darwinism has be studied intensively \cite{Riedel:2011,Riedel:2012,Zwolak:2013,Zwolak:2016}.

On the other hand, considering distinguishability between observables, which corresponds to orthogonality between states, SBS defines quantum state structure having objectivity of $\rho_{S:fE}$ is defined as an orthogonal convex combinatorial distribution of the environmental states $\rho_i$ over central observables $i$ which are uniquely selected by the total Hamiltonian and called pointer observables. Indeed, SBS implies the objectivity condition $I(\rho_{S:fE} )=H(\rho_S)$ but the
converse is still a open problem \cite{Le:2019mnn}. On SBS, the objective quantum state, which implies both perfect decoherence for a central observable $i$ and perfect distinguishability for the remaining environmental state $\rho^{E_k}_i$ after unobserved parts traced out, is realized as a total density matrix $\rho_{S:fE}$ as an orthogonal convex combination \cite{Korbicz:2014,Korbicz:2020tkf}:
\begin{align}\label{SBS}
\rho_{S:fE}=\sum_ip_i|i\rangle_S\langle i|\otimes\rho^{E_1}_i\otimes\cdots\otimes\rho^{E_{fN}}_i,
\end{align}
where 
\begin{align}\label{orthogality}   \rho^{E_k}_i\rho^{E_k}_j=0,~(i\neq j).
\end{align}
First, for given an arbitrary state $\rho_{S:fE}$, vanishing off-diagonal elements in $|i\rangle$ basis lead to a diagonal form in $|i\rangle$ basis. Tracing out ``unobserved'' part of the environmental fraction plays a main role in such a diagonal form through ``decoherence''. This numerical factor is called a ``decoherence factor'' \cite{Korbicz:2020tkf}. Second, orthogonality between associated states $\rho^{E_k}_i$ and $\rho^{E_k}_j$ for $i\neq j$ can be measured by a generalized overlap \cite{Korbicz:2020tkf}. The
smaller these factors get, the more distinguishable a central state is distributed.
A decoherence factor is responsible for how much a central observables are directly
distinguishable while a generalized overlap for how much a central observables are
distinguishable through environmental probe. The amount of distinguishability is accounted for the amount of information by entropy. Quantum Darwinism quantifies how many copies of the information of a central system is encoded into the environment as redundancy to achieve objectivity. 

Objectivity based on these mechanisms, has been studied in typical simple models, like a spin-spin model \cite{Zwolak:2016,Mironowicz:2017,Mironowicz:2018}, a spin-boson model \cite{Lampo:2017}, a boson-spin model \cite{Lee:2024iso,Lee:2024idx} and a boson-boson model \cite{Tuziemski:2015,Lee:2023ozm}. 

In practice, due to complexity of open quantum systems, their analysis has to rely on particular limits. A common approximation for open quantum systems is a weak coupling limit (Born approximation). In addition to it, it is often assumed that a central system is influenced by the environmental in a time local manner (no-memory or Markov approximation) while the environment remains almost undisturbed by central system. Such a combination of approximations is called, the Born-Markov approximation \cite{Schlosshauer2007}. 

However, our objectivity study here needs to allow the information of a central system to be encoded into the environment as the state of the environment changes. Thus, we introduce the appropriate approximation for the objectivity study, called the recoilless limit or the Born-Oppenheimer approximation \cite{Born:1927}. This approximation is that the evolution of a central system is effectively decoupled from the environment while the environment is influenced only by a classical trajectory of a central system.
The recoilless approximation can be theoretically justifiable by the robustness of objectivity for macroscopic objects \cite{Riedel:2011}. Once a macroscopic object achieves objectivity through decoherence, it does not go back to quantum superposition state, losing objectivity. Also, the recoilless limit simplify computation significantly.

The quantum Brownian motion (QBM) models have been one of major subjects in open quantum systems \cite{Ullersma:1966,Joos:2003,Petruccione:2010,Riedel:2011,Riedel:2012,Zwolak:2013,Zwolak:2016}. It is worthwhile to notice that objective quantum state structure in Eqs. \eqref{SBS} and \eqref{orthogality} is achieved only in an asymptotic sense as time goes infinity. Moreover, as we will see later, the asymptotic decay of the objectivity measures for quantum Brownian motion models with the finite number of the environments under the recoilless limit can occur only in a finite time range. The present work is motivated by the previous misinterpretation of the existence of objectivity without a relevant timescale in \cite{Tuziemski:2015}. 
The purpose of this article is to show how a general quantum state of a harmonic oscillator evolve a classical state with the recoilless assumption. Note we do not prove a general quantum state of a harmonic oscillator naturally evolves to a classical state without the recoilless assumption.
This article will show that objectivity in quantum Brownian motion models with the finite number of the environments exists relevant to time scale. We re-derive the objectivity measures for a general classical trajectory of a central oscillator, determine whether there exist non-objectivity conditions for infinite time domain and interpret at what timescales objectivity can be defined.

On the other hand, it has not been answered why a particular initial condition of a classical trajectory maximizes objectivity \cite{Paz:2009, Lee:2023ozm}. We will show explicitly why the phase $\phi=\pi/2$ in a classical trajectory $X_t=X_0\cos(\Omega t+\phi)$ maximizes objectivity.

 Although we consider only QBM for the existence of the objectivity time scale, we expect that this result can be valid for many models. Working for general results including redundancy of information in quantum Darwinism would be the next research topic.
 
 For the entire presentation we assume $\hbar=1$ and it should be clear from the context whether symbols are operators or numbers. 

%%%%%%%%%%%%%%%%%%%%%%%%%%%%%%%%%%%%%%%%%%%%%%%%%%%%%%%%%
  \section{Quantum Brownian motion}
The QBM model is a composite system of a central harmonic oscillator and an ensemble of the environmental harmonic oscillators with a position bi-linear interaction. $H$, the Hamiltonian for the QBM model \cite{Ullersma:1966,Joos:2003,Petruccione:2010}, is given by  
 \begin{align}
H&=H_{S}+\sum_kH^{(k)}_{E}+\sum_kH^{(k)}_{\text{int}},\label{QBM Hamiltonian}
 \end{align}
 where
\begin{align}\label{H_S,H_E,H_int}\nonumber
    H_{S}&=\frac{P^2}{2M}+\frac{1}{2}M\Omega^2 X^2,\\
    H^{(k)}_E&=\frac{p^2_k}{2m_k}+\frac{1}{2}m_k\omega^2_kx^2_k,\\
    H^{(k)}_{\text{int}}&= Xg_k x_k.\nonumber
 \end{align}
 Here $M$ and $\Omega$ are the mass and the angular frequency of a central oscillator, respectively and $m_k$ and $\omega_k$ are the mass and the angular frequency of the $k$th oscillator in the environment, respectively. $g_k$ is the coupling constant in a position bi-linear interaction. $H$ in Eq. \eqref{QBM Hamiltonian} identifies a central system through a bilinear interaction with the environmental oscillators. Also, we assume large mass $M\gg1$ compared to $g_k$. This may distinguish a central system from the environment with the opposite limit $m_k\ll1$ but here we do not necessarily impose $m_k\ll1$ for the environment. As in the next section we apply the recoilless limit, so-called Born-Oppenheimer approximation \cite{Born:1927} to the system to assume that a central system is not significantly influenced by interaction with the environment, large mass assumption $M\gg1$ with weak interaction $g\ll1$ will be the main assumption for Born-Oppenheimer approximation.
 
 In principle, the QBM Hamiltonian $H$ in Eq. \eqref{QBM Hamiltonian} can be written as two independent harmonic oscillators by choosing a linear combination of central and environmental basis. But the drawback of such a basis is that it loses original particle identities. Keeping the original particle identity basis can be regarded as the primary challenge in writing exact analytical solutions. Instead, keeping particle identities, we apply for an intermediate approximation, a so-called recoilless limit or the Born-Oppenheimer approximation \cite{Born:1927}. This approximation is based on the nature of a macroscopic object when the back-reaction of the environment on the central system is ignored and also in practice significantly reduces the complexity of the problem as the dynamics of a central system is decoupled and contributes to the environment as an external interaction with a classical trajectory of the central system.
 %%%%%%%%%%%%%%%%%%%%%%%%%%%%%%%%%%%%%%%%%%%%%%%%%%%%%%%
 \section{Recoilless limit}\label{Recoilless limit}
 It is difficult to express exact solutions for a unitary evolution operator for the QBM model in the original particle identity basis. The recoilless limit \cite{Born:1927} is a physically reasonable approximation for a macroscopic object and allows us to use the particle identity basis. Here we explain how the recoilless limit is obtained in the path integral formalism and the effective Hamiltonian $H_{\text{eff}}$ is obtained. The QBM model is a system of a driven central harmonic oscillator coupled with environmental oscillators. The kernel $\langle X_t|U_{S:E}| X_0\rangle\equiv K(X_t,t,X_0,0)$, i.e. a matrix element of a unitary evolution operator in the position basis between the final position $X_t$ at $s=t$ and the initial position $X_0$ at $s=0$, can be analytically expressed as \cite{Ingold:2002}, 
 \begin{align}\label{Kernel}
K(X_t,t,X_0,0)=\sqrt{\frac{M\Omega}{2\pi i\sin\Omega t}}e^{iS_{\text{cl}}[x(s)]},
 \end{align}
where 
\begin{align}\label{Scl}\nonumber
    &S_{\text{cl}}[x(s)]=\frac{M\Omega}{2\sin\Omega t}[(X^2_t+X^2_0)\cos\Omega t-2X_tX_0]\\
    &+\frac{g}{\sin\Omega t}\int^t_0ds x_k(s)\{X_t\sin\Omega s +X_0\sin[\Omega (t-s)]\}\\\nonumber
    &-\frac{g^2}{M\Omega\sin\Omega t}\int^t_0ds\int^s_0du\sin\Omega u\sin[\Omega (t-s)]x_k(u)x_k(s).
\end{align}
 In Eq. \eqref{Scl} the zeroth order in $S_{\text{cl}}$ is an action for a free harmonic oscillator and the first order is a position bi-linear interaction between the environmental $k$th oscillator and the classical trajectory of a central oscillator, parameterized by ($X_t$, $X_0$, $s$, $t$). If the second order contribution is negligible  (e.g., for small coupling \(g\ll 1\) and large mass \(M\gg 1\)), the total kernel $K_t\equiv \langle x_{kt},X_t|U_{S:E}|x_{k0}, X_0\rangle$, where $x_{kt}$ and $x_{k0}$ are the final and the initial positions of the environmental oscillator, respectively, is factored out as
 \begin{align}\label{Kt}
     K_t
     \approx\langle X_t|U_S|X_0\rangle \prod_k \langle x_{kt}|U^{(k)}_{\text{eff}}|x_{k0}\rangle.
 \end{align}
 The valid time scale for this truncation, $t_{re}$, is roughly estimated in the comparison between the first order and the second order terms:
\begin{align}\label{recoilless timescale}
gt_{re}\gg\frac{g^2t_{re}^2}{M\Omega}\to t_{re}\ll\frac{M\Omega}{g}. 
 \end{align}
The matrix elements $\langle X_t|U_S|X_0\rangle$ in Eq. \eqref{Kt}, i.e. the kernel for a free central oscillator is given by 
\begin{align}\label{S:kernel}\nonumber
     &\langle X_t|U_S|X_0\rangle\\\nonumber
     =&\int \mathcal{D}X\exp{i\int^t_0ds\left(\frac{P^2}{2M}-\frac{1}{2}M\Omega^2 X^2\right)}\\
=&\sqrt{\frac{M\Omega}{2\pi i\sin\Omega t}}e^{iS_0}
 \end{align}
 and the kernel $\langle x_{kt}|U_{\text{eff}}|x_{k0}\rangle$ for the $k$th environmental oscillator is
\begin{align}\label{Ueff:kernel}
    &\langle x_{kt}|U^{(k)}_{\text{eff}}|x_{k0}\rangle\\
    \nonumber
    =&\int \mathcal{D}x_k\exp{i\int^t_0ds\left[\frac{p^2_k}{2m_k}-\frac{1}{2}m_k\omega^2_i x^2_k+g_kX_{\text{cl}}(s)x_k \right]},
 \end{align}
 where $X_{\text{cl}}(s)$ from the first order in Eq. \eqref{Scl} is a general trajectory of a classical harmonic oscillator:
 \begin{align}\label{Xcl:X}
    X_{\text{cl}}(s)=\frac{1}{\sin\Omega t}(X_t-X_0\cos\Omega t)\sin\Omega s+X_0\cos\Omega s.
\end{align}
The operator $H_{\text{eff}}(s)$, where $U_{\text{eff}}\equiv e^{-i\int ^t_0dsH_{\text{eff}}(s)}$, can be read off from Eq. \eqref{Ueff:kernel} as
\begin{align}\label{Heff:Xcl}
    H_{\text{eff}}(s)=\frac{p^2_k}{2m_k}+\frac{1}{2}m_k\omega^2_i x^2_k-g_kX_{\text{cl}}(s)x_k.
\end{align}
It is not difficult to see that $X_{\text{cl}}(s)$ in Eq. \eqref{Xcl:X} is a general trajectory for a harmonic oscillator. Also, One notices that $X_{\text{cl}}(s)$ is written by four variables, a time variable $s$, a final time $t$, an initial position $X_0$ and a final position $X_t$, which are not all independent. For convenience and clarity that $X_{\text{cl}}(s)$ is indeed our familiar general trajectory, we introduce the $(Y,\phi)$ through $(X_0=Y\cos\phi,X_t=Y\cos(\Omega t+\phi))$, $X_{\text{cl}}(X_t,X_0;s,t)$.  
\begin{align}\label{Xcl:Y}\nonumber
    &X_{\text{cl}}(X_t,X_0;s,t)\\\nonumber
    &= \frac{1}{\sin\Omega t}[Y\cos(\Omega t+\phi)-Y\cos\phi\cos\Omega t]\sin\Omega s\\\nonumber
    &+\frac{1}{\sin\Omega t}Y\cos\phi\sin\Omega t\cos\Omega s\\
    &=-Y\sin\phi\sin\Omega s+Y\cos\phi\cos\Omega s\\\nonumber
    &=Y\cos(\Omega s+\phi).\nonumber
\end{align}
Since the difference between the variables $X_0$ and $Y$ is only a constant factor $\cos\phi$, the quantum state for the whole system $|\Psi_{S:E}\rangle$ can be expressed in either $|X_0\rangle $ or $|Y\rangle$ basis, using Eq. \eqref{Kt}: 
\begin{align}\label{effective state}\nonumber
 &|\Psi_{S:E}\rangle\\
 &=\int dX_0 e^{-i H_St}|X_0\rangle \langle X_0|\phi_0\rangle\otimes U_{\text{eff}}(X_0)|\psi_0\rangle\\
 &=\frac{1}{\cos\phi}\int dY e^{-i H_St}|Y\rangle \langle Y|\phi_0\rangle\otimes U_{\text{eff}}[Y\cos(\Omega t+\phi)]|\psi_0\rangle.\nonumber
\end{align}
Now since $X_{\text{cl}}$ is expressed by only the position variable  $Y$ and the time parameter $s$, instead of position $X_0$, the time $s$ and the final time $t$, it is clear to transform $H_{\text{eff}}(s)$ in Eq. \eqref{Heff:Xcl} into an operator form. Putting Eq. \eqref{S:kernel} and Eq. \eqref{Ueff:kernel} together into Eq. \eqref{Kt}, one can see that the approximate expression of $K_t$ in Eq. \eqref{Kt} is the ansatz proposed in Refs. \cite{Tuziemski:2015,Lee:2023ozm}, which is based on the dominance of a central system energy-scale over the environment and suppresses the environmental back-reaction to a central system:
\begin{align}\label{Kt:ansatz}\nonumber
     K_t&\approx \sqrt{\frac{M\Omega}{2\pi i\sin\Omega t}}e^{iS_{\text{sys}}[X_{\text{cl}}(t)]} \\
&\times\int\mathcal{D}x(t)e^{iS_{\text{env}}[x(t)]}e^{iS_{\text{int}}[X_{\text{cl}} (t),x(t)]}.
 \end{align}  
%%%%%%%%%%%%%%%%%%%%%%%%%%%%%%%%%%%%%%%%%%%%%%%%%%%%%%%%%%%%
\section{Exact solutions}
The effective Hamiltonian $H_{\text{eff}}$ from Eq. \eqref{Heff:Xcl} and Eq. \eqref{Xcl:Y} is expressed as
\begin{align}\label{Heff:Y}
 H^{(k)}_{\text{eff}}(t)=\frac{p^2_k}{2m_k}+\frac{1}{2}m_k\omega^2_kx^2_k-Yg_kx_k\cos(\Omega t+\phi).
 \end{align}
Now, our task is reduced to solving the environmental driven oscillators coupled with the corresponding classical trajectory of a central oscillator, i.e. the periodic Hamiltonian $H^{(k)}_{\text{eff}}$ with the coupling of the classical trajectory $Y\cos(\Omega t+\phi)$. 

Here, for the objectivity analysis we will especially use the Floquet theory, which is known as a theorem for a periodic system \cite{Floquet:1883}. For a periodic Hamiltonian $H^{(k)}_{\text{eff}}$ in Eq. \eqref{Heff:Y}, the Floquet theorem \cite{Floquet:1883} expresses the unitary evolution operator $U^{(k)}_{\text{eff}}$ as a product of periodic and non-periodic unitary operators, in general:
\begin{align}\label{Ueff}
U^{(k)}_{\text{eff}}=e^{-iK^{(k)}_Y}e^{-iH^{(k)}_{F}t}e^{iK^{(k)}_{Y0}},
 \end{align}
 where $K^{(k)}_Y$ is a periodic time-dependent Hamiltonian associated with a classical trajectory $Y\cos(\Omega t+\phi)$. This separation property between the periodic part and the rest is especially important in  the objectivity analysis for periodic systems. Since the objectivity for a periodic system requires non-periodicity, periodic parts from a unitary evolution needs to be separated. 

Fortunately, the exact analytical solution for $U^{(k)}_{\text{eff}}$ is known in \cite{Husimi:1953, Tuziemski:2015,Lee:2023ozm, Hanggi:2016, Bukov:2017}. The Floquet Hamiltonian $H^{(k)}_{F}$ in Eq. \eqref{Ueff} is a harmonic oscillator Hamiltonian up to a constant. It is important to identify which part is relevant to objectivity. If the periodicity is preserved in the objectivity measures, it cannot be expected to have objectivity. The Floquet method can show which conditions break a periodicity. For the QBM model, the exact solution for $U^{(k)}_{\text{eff}}$ in Eq. \eqref{Ueff} can be written in the Floquet composition up to a global phase $e^{i\delta_k}$ \cite{ Hanggi:2016,Bukov:2017}:
 \begin{align}\label{Ueff:final}
U^{(k)}_{\text{eff}}&=e^{-iK^{(k)}_Y}e^{-iH^{(k)}_{F}t}e^{iK^{(k)}_{Y0}},
\end{align}
where
\begin{align}\label{KzetaHF}\nonumber
     e^{-iK^{(k)}_Y}&=e^{i\delta_k}e^{-i\zeta^{(k)}_Y p_k+im_k\partial_t\zeta^{(k)}_Y x_k},\\
     \zeta^{(k)}_Y&=-Yg_kR_k\cos(\Omega t+\phi),\\
H^{(k)}_F&=H^{(k)}_0+\frac{1}{4}Y^2g^2_kR_k,\nonumber
 \end{align}
 with
 \begin{align}\label{Rk}
     R_k\equiv\frac{1}{m_k(\omega^2_k-\Omega^2)}.
 \end{align}
Here, $-iK^{(k)}_Y$ is more explicitly expressed as
 \begin{align}\label{KY}\nonumber
     -iK^{(k)}_Y&=iYg_kR_k[cp_k-m\Omega sx_k]\\
       &=-Y[\alpha_k a^\dagger_k-\alpha^*_k a_k],
 \end{align}
 %&=iYA\left[ic\sqrt{\frac{m\omega}{2}}(a^\dagger-a)-\frac{m\Omega s}{\sqrt{2m\omega}}(a^\dagger+a)\right]\\\nonumber
      % &=YA\sqrt{\frac{m\omega}{2}}\left[\left(c-i\frac{\Omega}{\omega}s\right)a-\left(c+i\frac{\Omega}{\omega}s\right)a^\dagger\right]\\
 where $c\equiv\cos(\Omega t+\phi)$, $s\equiv\sin(\Omega t+\phi)$, $(a_k,a^\dagger_k)$ are a creation operator and an annihilation operator for a harmonic oscillator, respectively and
 \begin{align}\label{alpha}
    \alpha_k\equiv R_k\sqrt{\frac{m_k\omega_k}{2}}\left(c+i\frac{\Omega}{\omega_k}s\right).
 \end{align}
 The periodic Floquet unitary evolution $e^{-iK^{(k)}_Y}$ is the displacement operator defined by  $D(\alpha)\equiv e^{\alpha a^\dagger-\alpha^* a}$, where the parameter $\alpha$ is given by $-Y\alpha_k$:
 \begin{align}\label{K_displacement}
     e^{-iK^{(k)}_Y}=D^\dagger[Y\alpha_k]=D[-Y\alpha_k].
 \end{align}
 %%%%%%%%%%%%%%%%%%%%%%%%%%%%%%%%%%%%%%%%%%%%%%
\section{Objectivity markers}
The spectrum broadcast structures (SBS) in Eq. \eqref{SBS} and Eq. \eqref{orthogality} have two physical meanings, decoherence for a central system and distinguishability for the environmental state. If pointer states for a central system are uniquely chosen by interaction with environment through a perfect decoherence process, pointer states should be perfectly distinguishable, i.e. in a convex combinatorial form with the corresponding associated environmental states on a density matrix. In this process, pointer state information on a central system is encoded into the environment. On the other hand, when an observer takes a measurement on the environmental system, distinguishability between environmental states corresponding to pointer states is relevant for obtaining the encoded information. These two physical concepts can be characterized by two objectivity markers, a decoherence factor $|\Gamma_{Y,Y'}|^2$ and a generalized overlap $B_{Y,Y'}$ \cite{Helstrom:1969}. Their asymptotic decay in time, $|\Gamma_{Y,Y'}|^2\to0$ and $B_{Y,Y'}\to0$ will confirm objectivity. In quantum Darwinism, SBS structure implies that the amount of information encoded from a central system $S$ into the environmental macro fraction $fE$, quantum mutual information is equal to the information of $S$, i.e. $I(\rho_{S:fE})=H(\rho_S)$, where $H(\rho_S)$ is the von Neumann entropy of $S$. Therefore, when both a decoherence factor and a generalized overlap decay in time implies that the information of a central system is encoded into a fractional environment $fE$. Here we are more interested in a generalized overlap for its analytical property for distinguishability measure in time, rather than quantum Chernoff bound, a measure of asymptotic behaviour of distinguishability between two large number of copies of states \cite{Zwolak:2016}.

 Using the factorization property of the kernel that we found in Eq. \eqref{Kt}, the matrix element of the total density matrix $\rho_{S:E}$ only for a central system, $\langle Y'|\rho_{S:E}|̺Y\rangle$ is written as
\begin{align}\label{partial matrix element}
&\langle Y'|\rho_{S:E}|̺Y\rangle\\
\approx&\int dYdY'
\langle Y'|\rho_{S0}|̺Y\rangle K_SK_S^*\bigotimes_kU^{(\text{k})}_{\text{eff}}[X_{\text{cl}}]\rho^{k}_{E0}U^{(\text{k})\dagger}_{\text{eff}}[X_{\text{cl}}].\nonumber
\end{align}
Assuming separate initial environmental states, after the unobserved environmental degrees of freedom traced out, $\langle Y'|\rho_{S:E}|̺Y\rangle$ is written as 
    \begin{align}\label{partial matrix element:unob removed}\nonumber
&\langle Y'|\rho_{S:E_{\text{obs}}}|̺Y\rangle\\\nonumber
\approx&\int dYdY'
\langle Y'|\rho_{S0}|̺Y\rangle K_SK_S^*F[X_{\text{cl}},X'_{\text{cl}}]\\
&\bigotimes_{k\in E_{\text{obs}}}U^{(\text{k})}_{\text{eff}}[X_{\text{cl}}]\rho^{k}_{E0}U^{(\text{k})\dagger}_{\text{eff}}[X_{\text{cl}}],
\end{align}
where $K_S=\langle X_t|U_S|X_0\rangle$ and $F[X_{\text{cl}},X'_{\text{cl}}]$ is the part traced out defined by
\begin{align}\label{F}
F[X_{\text{cl}},X'_{\text{cl}}]&\equiv\prod_{{k\in E_{\text{uno}}}}\text{Tr}\{U^{(\text{k})}_{\text{eff}}[X_{\text{cl}}]\rho^{k}_{E0}U^{(\text{k})\dagger}_{\text{eff}}[X_{\text{cl}}]\}.
\end{align}
According to the SBS structure Eq. \eqref{SBS} and Eq. \eqref{orthogality}, the objectivity for QBM can be characterized by two properties, the orthogonal structures between pointer states $\{|Y\rangle\}$ and between the corresponding associated environmental states $\{\rho^{E_k}_Y\}$ in $\rho_{S:E_{\text{obs}}}$.

First, we see that for  $Y\neq Y'$, if the matrix element of $\rho_{S:E_{\text{obs}}}$ in Eq. \eqref{partial matrix element:unob removed} vanishes, i.e. $\langle Y'|\rho_{S:E_{\text{obs}}}|̺Y\rangle=0$ , it guarantees the orthogonal convex combinatorial distribution for pointer variables $\{Y\}$ in $\rho_{S:E_{\text{obs}}}$. The numerical factor $F[X_{\text{cl}},X'_{\text{cl}}]$ in Eq. \eqref{F} called a decoherence factor is responsible for a degree of decoherence and such orthogonal distribution of pointer states as well.

Second, for objectivity it is also necessary to characterize the environmental orthogonal structure, $\rho^{E_k}_Y\rho^{E_k}_{Y'}=0$ for $Y\neq Y'$. The numerical factor responsible for the associated environmental orthogonality with respect to a pointer state, called a generalized overlap (fidelity), is defined below in Eq. \eqref{objectivity markers}.

On the other hand, SBS can be viewed on quantum Darwinism. It is known that if $\rho_{S:E_{\text{obs}}}=\sum_{i}p_i|i\rangle\langle i|\otimes\rho_ i$, the mutual information $I(\rho_{S:E_{\text{obs}}})=H(\sum_ip_i\rho_i)-\sum_ip_iH(\rho_i)$. If $\rho_i\rho_{j}=0$ for $i\neq j$, by changing basis of $\rho_i$ to $|i\rangle\langle i|$, $I(\rho_{S:E_{\text{obs}}})=H(p_i|i\rangle\langle i|)=H(S)$   $I(\rho_{S:E_{\text{obs}}})=H(\rho_S)$, which implies that the common information between a central system and the environment is the same as one for a central central system.  as explained in Introduction \ref{Introduction}.

Ignoring phase factors, these two measures for objectivity are defined by
%1)The authors should clearly explain why the objectivity markers are defined as in Eq.(27). What are the physical meanings of the two quantities, i.e., the decoherence factor and distinguishability? In particular, the authors should explain, from the perspective of quantum Darwinism, why these quantities serve as indicators of objectivity and what physical questions they are intended to answer.

\begin{align}\label{objectivity markers}
    |\Gamma_{Y,Y'}|^2&=\left|\text{Tr}[U_{Y}\rho_0U^\dagger_{Y'}]\right|^2,\\
    B_{Y,Y'}&=\left[\text{Tr}\sqrt{\sqrt{U_{Y'}\rho_0 U^\dagger_{Y'}}U_{Y}\rho_0 U^\dagger_{Y}\sqrt{U_{Y'}\rho_0 U^\dagger_{Y'}}}\right]^2.\nonumber
\end{align}
In our case $U_Y\equiv \prod\limits_{k} U^{(k)}_{\text{eff}}$, where $U^{(k)}_{\text{eff}}$ is given in Eq. \eqref{Ueff}, and $\rho_0$ is assumed to be an initial separate density matrix for the environment:
\begin{align}\label{intial separate state}
    \rho_0=\prod_k\rho^{(k)}_0.
\end{align}
When these quantities vanish, it confirms the objective structure.
We list general properties of the objectivity markers as follows. 
\begin{enumerate}
    \item \textit{Domain}: Objectivity markers take values between 0 and 1.
    \begin{align}\label{objectivity ranges}
        0\leq|\Gamma_{Y,Y'}|^2\leq1,~0\leq B_{Y,Y'}\leq1.
    \end{align} 
    \item \textit{Initial value}: Initial values for the objectivity markers should be the maximum value, 1.
    \begin{align}\label{initial objectivity}
        |\Gamma_{Y,Y',0}|^2=B_{Y,Y',0}=1.
    \end{align}
    \item \textit{Structure}: $|\Gamma_{Y,Y'}|^2$ is asymmetric about $\rho_0$, i.e. a function of $U_Y\rho_0U^\dagger_{Y'}$ while $B_{Y,Y'}$ is symmetric about $\rho_0$, i.e. a function of $U_Y\rho_0U^\dagger_{Y}$ and $U_{Y'}\rho_0U^\dagger_{Y'}$.
    \item \textit{Multiplicativity}: In the absence of mutual interactions between environments, the objectivity markers can be written a as a product of the objectivity markers for individual environmental systems.
    \begin{align}\label{objectivity markers prod}\nonumber
       |\Gamma_{Y,Y}|^2&=\prod_k |\Gamma^{(k)}_{Y,Y}|^2,\\
       B_{Y,Y'}&=\prod_k B^{(k)}_{Y,Y'}.
    \end{align}
    This multiplicative property and the domain of the objectivity markers imply that as the number of environments increased, the objectivity markers at any time get smaller as much as possible, unless the values of the individual objectivity markers are exactly unity. In periodic systems with the multiplicativity there may exist such non-objectivity points.
    \item Due to our definitions for objectivity markers Eq. \eqref{objectivity markers} any overall phases do not contribute to objectivity.
\end{enumerate}
\section{Objectivity conditions}
Objectivity is determined by both an asymptotically decaying decoherence factor and an asymptotically decaying generalized overlap in time. Here we consider the general objectivity conditions for the QBM model. 
\begin{enumerate}
    \item A multi-environmental system is the most necessary ingredient for asymptotically decaying objectivity markers. Since total objectivity markers for non-mutually interacting environments are simply a product of the individual objectivity makers as shown in Eq. \eqref{objectivity markers prod} they can get smaller and smaller as more environmental systems are added, if there exist no special maximum points. It is worth checking such 
 points, especially in periodic environments.
    \item It is important to check cases in which the maximum values of objectivity markers for the environments, i.e. ``one'' in Eq. \eqref{objectivity ranges} appear repeatedly at the same time spots as time goes on. Moreover, when such time points from each environment systems coincide with each other as time goes to infinity, any decay in the objectivity markers will not occur. As will be shown in the next section, there exist conditions for frequencies making such maxima occur periodically. The origin of such a recurrence is the common frequency from a central system transferred to the environment. Breaking such a periodicity in the environment leads to objectivity. For the QBM model under the recoilless limit, the effective Hamiltonians for environmental oscillators obtain the common frequency from a central oscillator. In this case it is important to examine whether a particular common frequency exists in the objectivity markers.
    \item The structural difference between a decoherence factor and a generalized overlap may provide answer to the question why distinguishability can be harder to obtain than decoherence, or why the distinguishability length scale is longer than the decoherence length scale \cite{Lee:2024iso,Lee:2023ozm}.  For a periodic Hamiltonian for the environment with a common frequency over all the environmental systems, the Floquet theorem separates a unitary evolution responsible for the periodicity from the rest. The relevant operator $B$ for a generalized overlap $B_{Y,Y'}=|\text{Tr}\sqrt{B}|^2$ in Eq. \eqref{objectivity markers} defined by
    \begin{align}\label{B}
        B\equiv\sqrt{\rho_0}U_Y^\dagger U_{Y'}\rho_0 U^\dagger_{Y'}U_Y\sqrt{\rho_0},
    \end{align}
     where $U_Y=e^{-iK_Y}e^{-iH_F(Y)t}e^{iK_{Y0}}$. 
     The operator $K_Y$ is periodic with a common frequency $\Omega$ and $H_F$ is not periodic, in general. One can see that $H_F$ plays a key role breaking the common periodicity. Due to the symmetric structure of $B$, for instance if 
     \begin{align}\label{non-distinguishability condition}
         [e^{-iH_F t}e^{iK^{(k)}_{Y0}},\rho^{(k)}_0]=0,
     \end{align}
    the contribution $H_F(Y)$ is canceled out and only the periodicity from $K_Y$ becomes a periodicity in $\text{Tr}\sqrt{B^{(k)}}$ and the initial maximum value ``one'' in a generalized overlap will appear periodically without any decay, which applies to all the environmental systems. Thus, a total generalized overlap in this case is periodic without a decay. On the other hand, since the operator $A$ for a decoherence factor is asymmetric around $\rho_0$, i.e.
      \begin{align}\label{A}
    A&\equiv U_Y\rho_0U^\dagger_{Y'},
    \end{align}
    it is more difficult to find an operational relation like Eq. \eqref{non-distinguishability condition} for non-objectivity in a generalized overlap to eliminate a periodicity-breaking contribution from $H_F$.
\end{enumerate}
%%%%%%%%%%%%%%%%%%%%%%%%%%%%%%%%%%%%%%%%%%%%%%%%%%%%%%%%%
\section{Objectivity in QBM}
Our main purpose in this section is to revisit the objectivity conditions for the QBM model under the recoilless limit with an initial thermal state for $\rho_0=\bigotimes_k\rho^{(k)}_{\text{th}}$ previously studied in \cite{Tuziemski:2015, Lee:2023ozm}. The thermal state $\rho^{(k)}_{\text{th}}$ for the $k$th environment is a function of the Hamiltonian for a harmonic oscillator, $H^{(k)}_0$,
\begin{align}
    \rho^{(k)}_{\text{th}}=\frac{e^{-\beta H^{(k)}_0}}{\text{Tr}[e^{-\beta H^{(k)}_0}]},
\end{align}
where $\beta\equiv1/k_BT$, $k_B$ is the Boltzmann constant and $T$ is temperature.
    Also, $H_F$ turns out to be the Hamiltonian for a harmonic oscillator up to a constant. Equivalently, Ref. \cite{Lee:2023ozm} computes $U^{(k)}_{\text{eff}}$ using the interaction picture absorbing the Hamiltonian for a harmonic oscillator. In particular, the QBM structure under the recoilless limit with a thermal initial state allows us to obtain the exact objectivity markers without any further approximation. Refs. \cite{Tuziemski:2015, Lee:2023ozm} derive exact objectivity markers in this setting with particular initial classical trajectories, i.e. at $\phi=0$ and $\phi=\pi/2$. This section analyzes the objectivity markers with a general phase $0\leq\phi\leq \pi/2$, which are derived in Appendix \ref{appendix}. One can see that the objectivity markers obtained here at $\phi=0$ and $\phi=\pi/2$ are the same as those obtained in \cite{Tuziemski:2015, Lee:2023ozm}. A decoherence factor and a generalized overlap for QBM in Eq. \eqref{A:decoherence factor} and Eq. \eqref{A:generalized overlap}, respectively, are derived as
\begin{align}\label{decoherence factor}
     |\Gamma_{Y,Y'}|^2
     &=\prod_k e^{-\coth\left(\frac{\beta\omega_k}{2}\right)(Y-Y')^2|\eta_k|^2}
\end{align}
and 
\begin{align}\label{generalized overlap}
    B_{Y,Y'}=\prod_ke^{-\tanh\left(\frac{\beta\omega_k}{2}\right)(Y-Y')^2|\eta_k|^2},
\end{align}
where 
\begin{align}\label{eta}
    \eta_k&=g_kQ_k\left[-e^{-i\omega_k t}\left(c+i\frac{\Omega}{\omega_k}s\right)+c_0+i\frac{\Omega}{\omega_k}s_0\right]
\end{align}
with the notations $c\equiv\cos(\Omega t+\phi)$, $s\equiv \sin(\Omega t+\phi)$, $c_0\equiv\cos\phi$ and $s_0=\sin\phi$ have been used and the resonant factor $Q_k$ 
\begin{align}\label{Q}
    Q_k\equiv\sqrt{\frac{\omega_k}{2m_k(\omega^2_k-\Omega^2)^2}}.
\end{align}

Temperature and the interaction coupling play a role only in enhancing and reducing objectivity without determining it. As found in \cite{Tuziemski:2015, Lee:2023ozm}, the coefficients $\coth(\beta\omega_k/2)$ and $\tanh(\beta\omega_k/2)$ in Eq. \eqref{decoherence factor} and Eq. \eqref{generalized overlap} indicate that at high temperature, $\beta\to 0$, a decoherence factor tends to get smaller while a generalized overlap becomes larger. On the other hand, at lower temperature, $\beta\to\infty$, the difference of the temperature dependence between a decoherence factor and a generalized overlap gets smaller.
As seen in Eq. \eqref{eta} the stronger the $g_k$, 
the stronger objectivity.
%%%%%%%%%%%%%%%%%%%%%%%%%%%%%%%%%%%%%%%%%
\subsection{Beating effect}
\begin{figure}[t!]
\includegraphics[width=1.0\linewidth, height=4cm]{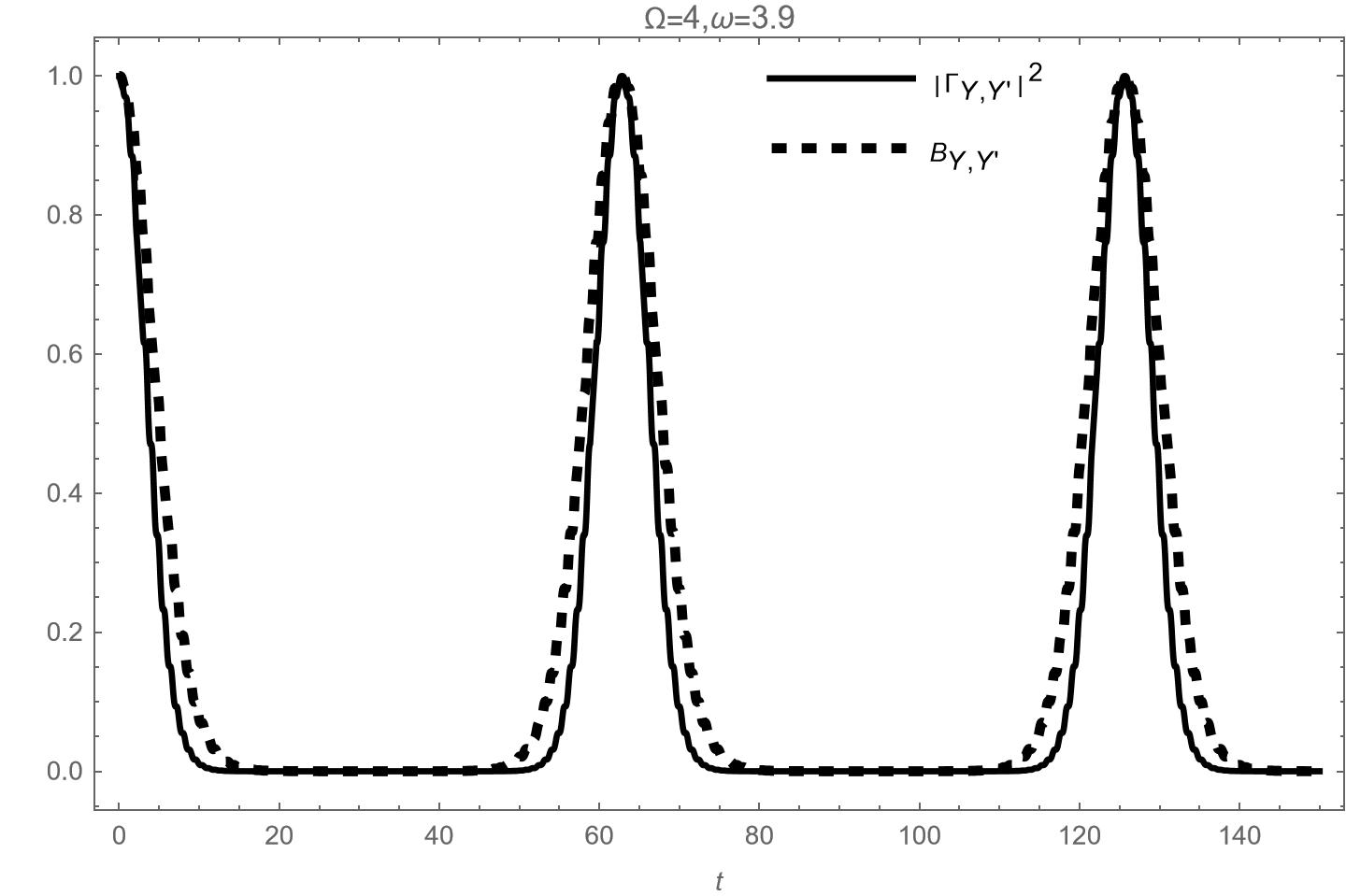} 
\caption{\label{fig:Resonance} The objectivity markers at $(\Omega,\omega)=(4,3.9)$ form a large profile by a beating frequency $\Delta\omega=\Omega-\omega=0.1$.}
\end{figure}

In the limit $\omega_k \to\Omega$, $Q_k$ in Eq. \eqref{Q} diverges but $\eta$ in Eq. \eqref{eta} is finite as shown in  Appendix \ref{appendix C}, so that there is no resonance issue. Nevertheless, as shown in Eq. \eqref{A:eta bar absolute cosines}, $\bar{\eta}_k\equiv\eta_k/g_kQ_k$ is expressed as a linear combination of cosines with frequencies $\Omega+\omega_k$, $2\Omega$ and $\Omega-\omega_k$ apart from a constant term:
\begin{align}\label{eta bar absolute cosines}\nonumber
|\bar{\eta}_k|^2
 &=-\frac{1}{2}\left(1+\frac{\Omega}{\omega_k}\right)^2\cos(\Omega-\omega_k)t\\\nonumber
 &-\frac{1}{2}\left(1-\frac{\Omega^2}{\omega^2_k}\right)\cos[(\Omega-\omega_k)t+2\phi]\\\nonumber
 &-\frac{1}{2}\left(1-\frac{\Omega}{\omega_k}\right)^2\cos(\Omega+\omega_k)t\\
 &-\frac{1}{2}\left(1-\frac{\Omega^2}{\omega^2_k}\right)\cos[(\Omega+\omega_k)t+2\phi]\\
 &+1+\frac{\Omega^2}{\omega^2_k}+\frac{1}{2}\left(1-\frac{\Omega^2}{\omega^2_k}\right)[\cos2(\Omega t+\phi)+\cos2\phi].\nonumber
\end{align}
In Eq. \eqref{eta bar absolute cosines}, $\omega_k\simeq\Omega$ forms a large envelop with the frequency $\Delta\omega_k=|\Omega-\omega_k|$ while the rest frequencies form smaller fluctuations on the large envelope. The emergent small frequency $\Delta\omega_k$ due to the environmental Hamiltonian plays a role breaking the periodicity of a central frequency $\Omega$, which is essential in objectivity of the boson-spin model \cite{Lee:2024iso,Lee:2024idx}. Such a small frequency makes a large timescale. Although the objectivity makers have a periodicity with such a large timescale, as the environment consists of more oscillators creating beatings with different frequencies, the periodicity and the timescale become larger and larger. Fig. \ref{fig:Resonance} shows that the objectivity markers at $\omega_k\simeq\Omega$ form a large profile. For $\omega_k\simeq\Omega$ the phase $\phi$ effect is small. As shown in Appendix \ref{A:Beating effect}, $\phi$ influences $|\bar{\eta_k}|^2$ such that for $\Omega>\omega_k$, $|\bar{\eta_k}|^2$ increases as $\phi$ increases; conversely, for $\Omega<\omega_k$, it decreases. $\phi$ affects the objectivity in the same way as the general phase relation which will be found in the section \ref{Classical trajectory}. 
%%%%%%%%%%%%%%%%%%%%%%%%%%%%%%%%%%%%%%%%%%%%%%%%%%%%%%%
%%%%%%%%%%%%%%%%%%%%%%%%%%%%%%%%%%%%%%%%%%%%%%%%%%%%%%%
%%%%%%%%%%%%%%%%%%%%%%%%%%%%%%%%%%%%%%%%%%%%%%%%%%%%%%%%%%%%%%%%%%%%%
\subsection{Fractional relation in frequency}
\begin{figure}[t!]
\includegraphics[width=1.0\linewidth, height=4cm]{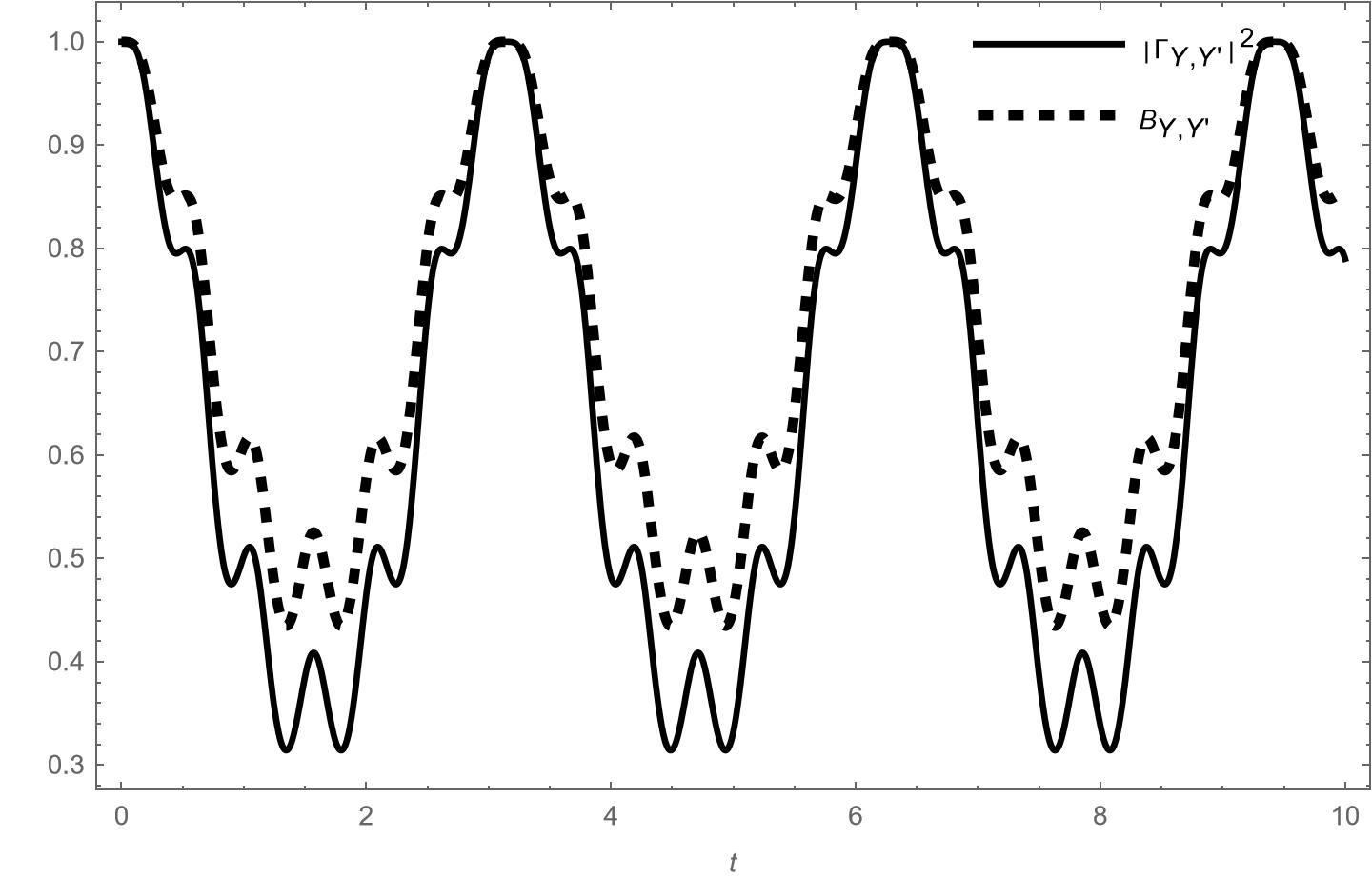} 
\caption{\label{fig:Omega7omega5} The objectivity markers with $\omega=5$ and $\Omega=7$. $t^o_{\text{min}}=\pi$. $t_{n}=\pi p$, ($p=1,2,3,\cdots$). }
\end{figure}

One may think that the exponential forms of the objectivity markers in Eq. \eqref{decoherence factor} and Eq. \eqref{generalized overlap} always make the objectivity markers get closer to 0 as more systems are added to the environment. 
However, if there exist an infinite series of time points where $\eta_k=0$, which causes the amplitudes of objectivity markers to be unity and coincide for all the environments, the multiplicativity property of the objectivity markers will not contribute to reducing such amplitudes at those points. 
Given the objectivity markers Eq. \eqref{decoherence factor} and Eq. \eqref{generalized overlap}, the two types of solutions of $\omega_k$ for $\eta_k=0$ are found in Appendix \ref{appendix C}:
\begin{align}\label{even condition}
 \frac{2n\pi}{\Omega}&=\frac{2m\pi}{\omega_k}=t_{2n},~(n.m\in \mathbb{N})
\end{align}
and
\begin{align}\label{odd condition}
\frac{(2n+1)\pi}{\Omega}&=\frac{(2m+1)\pi}{\omega_k}=t_{2n+1},~(n.m\in \mathbb{N}).
\end{align}
These conditions Eq. \eqref{even condition} and Eq. \eqref{odd condition} constrain the frequency ratio $\frac{\omega_k}{\Omega}$ by a reduced fraction of a frequency ratio $f_{\text{re}}=\frac{\omega_k}{\Omega}$ and specify the associated non-objectivity times points, $t_{2n}$ and $t_{2n+1}$. The reduced fraction $f_{\text{re}}$ can be one of the forms $f_{\text{re}}=\frac{\text{odd}}{\text{even}}, \frac{\text{even}}{\text{odd}}$ and $\frac{\text{odd}}{\text{odd}} $. $f^e_{\text{re}}=\frac{\text{odd}}{\text{even}}, \frac{\text{even}}{\text{odd}}$ correspond to the condition Eq. \eqref{even condition} and $f^o_{\text{re}}=\frac{\text{odd}}{\text{odd}}$ can correspond to both Eq. \eqref{even condition} and Eq. \eqref{odd condition} but only Eq. \eqref{odd condition} is sufficient.  The earliest time $t^e_{\text{min}}$ for $\eta_k=0$ after $t=0$, by satisfying the conditions Eq. \eqref{even condition} and Eq. \eqref{odd condition}, is
\begin{align}\label{earliest even time}
    t^e_{\text{min}}&=\frac{2n_{\text{min}}}{\Omega}\pi,~(n_{\text{min}}\in\mathbb{N})
\end{align}
and one for $f^o_{\text{re}}$, $t^o_{\text{min}}$ is
\begin{align}\label{earliest odd time}
     t^o_{\text{min}}&=\frac{2n_{\text{min}}+1}{\Omega}\pi,~(n_{\text{min}}\in\mathbb{N}).
\end{align}
The non-objectivity time points $\{t_M\}$ at which the objectivity makers become one are multiples of $t_{\text{min}}$: 
\begin{align}\label{non-objevtivity odd even sequences}\nonumber
   t^e_M&=pt^e_{\text{min}},~(p\in\mathbb{N}),\\
   t^o_M&=pt^o_{\text{min}},~(p\in\mathbb{N}).
\end{align}
Fig. \ref{fig:Omega7omega5} illustrates the fractional relation with the example $(\Omega,\omega_k)=(7,5)$ corresponding to either of the conditions Eq. \eqref{even condition} and Eq. \eqref{odd condition} where the objectivity markers are first maximized at $t_{\text{min}}=\pi$.
%%%%%%%%%%%%%%%%%%%%%%%%%%%%%%%%%%%%%%%%%%%%%%%%%%%%%%%
%%%%%%%%%%%%%%%%%%%%%%%%%%%%%%%%%%%%%%%%%%%%%%%%%%%%%%%%%%

To determine the non-objectivity time sequence, both a given ratio $\frac{\omega_k}{\Omega}$ and the value of either $\Omega$ or $\omega_k$ are required. This means that for a given $\Omega$ there exists a set $\{\omega_p\}$ having the same $t_{\text{min}}$ and hence $t_M$:
\begin{align}\label{even frequency set}
    \omega^e_{p,n_{\text{min}}}=\frac{2p}{2n_{\text{min}}}\Omega,~(p\in\mathbb{N},p\neq0)
\end{align}
and
\begin{align}\label{odd frequency set}
    \omega^o_{p,n_{\text{min}}}=\frac{2p+1}{2n_{\text{min}}+1}\Omega,~(p\in\mathbb{N}).
\end{align}
The comparison between the cases for the same ratio between $\omega_k$ and $\Omega$ but having different sets of $t_{\text{min}}$ are shown in Fig. \ref{fig:Omega3omega2,Omega6omega4}. If the systems of the environment belong to Eq. \eqref{even frequency set} or Eq. \eqref{odd frequency set}, all the non-objectivity points have the amplitude 1 in the objectivity markers and there is no objectivity. 

One may think that if the set $\{\omega_k\}$ in the environment has different non-objectivity sequences, the non-objectivity sequences could be canceled out. However, there always exists a common infinite sequence of the non-objectivity points according to Eq. \eqref{non-objevtivity odd even sequences}. Therefore, objectivity cannot be achieved with the fractional frequency relation. It implies that if environmental oscillators with frequencies having the non-fractional relation with $\Omega$ are added, the periodic pattern can be destroyed and the objectivity markers can get closer to 0. In a practical sense, if the environmental frequencies follows fractional relation approximately, periodic non-objectivity will be well maintained approximately.  
%%%%%%%%%%%%%%%%%%%%%%%%%%%%%%%%%%%%%%%%%%%%%%%%%%%%%%%%%
%%%%%%%%%%%%%%%%%%%%%%%%%%%%%%%%%%%%%%%%%%%%%%%%%%%%%%%
\begin{figure}[t!]
\begin{subfigure}{.2\textwidth}
  \includegraphics[width=1.1\linewidth]{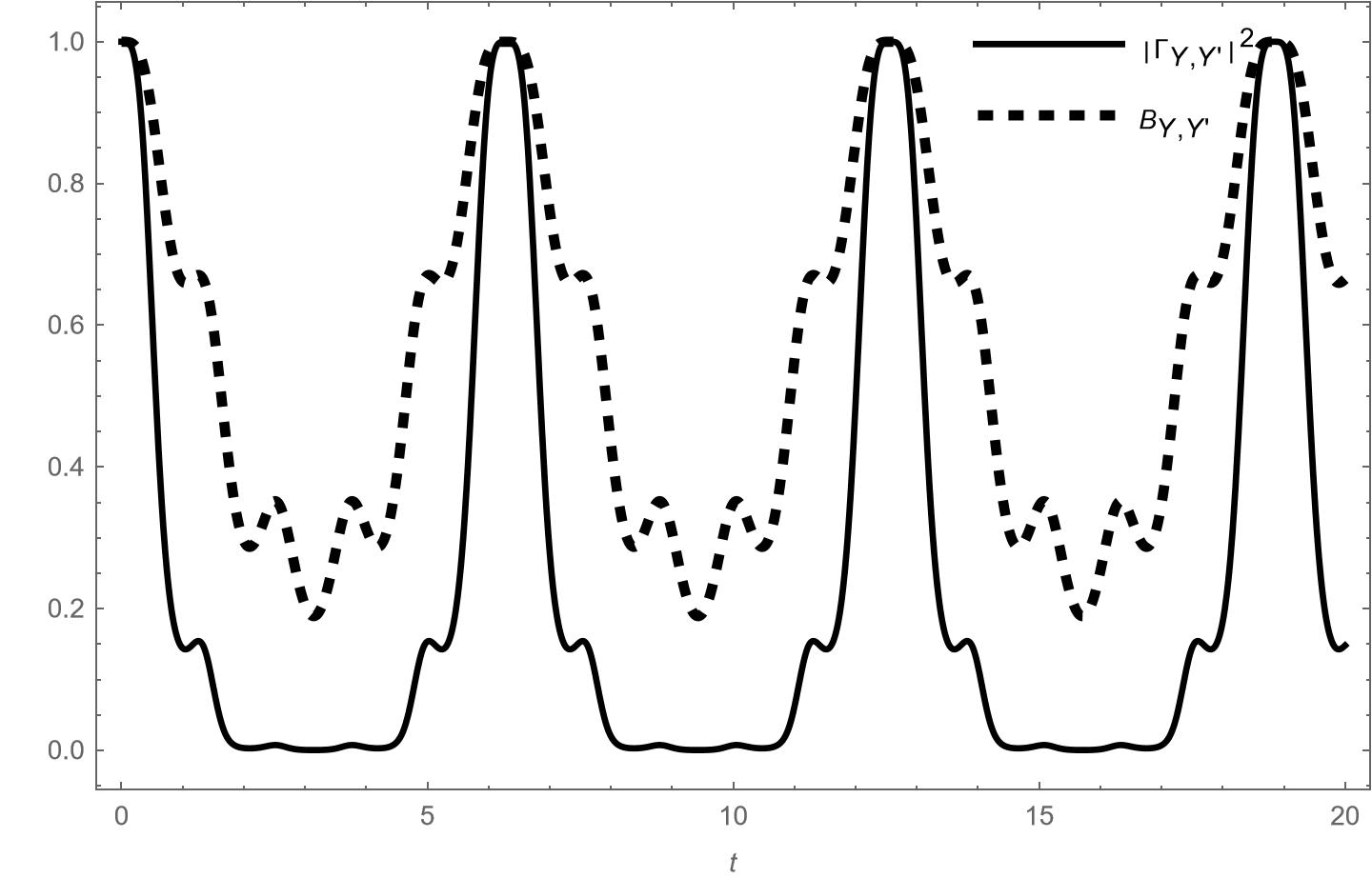}
  \caption{$\omega=2$ and $\Omega=3$. $t^e_{\text{min}}=2\pi$.}
\end{subfigure}%
\begin{subfigure}{.2\textwidth}
  %\centering
  \includegraphics[width=1.1\linewidth]{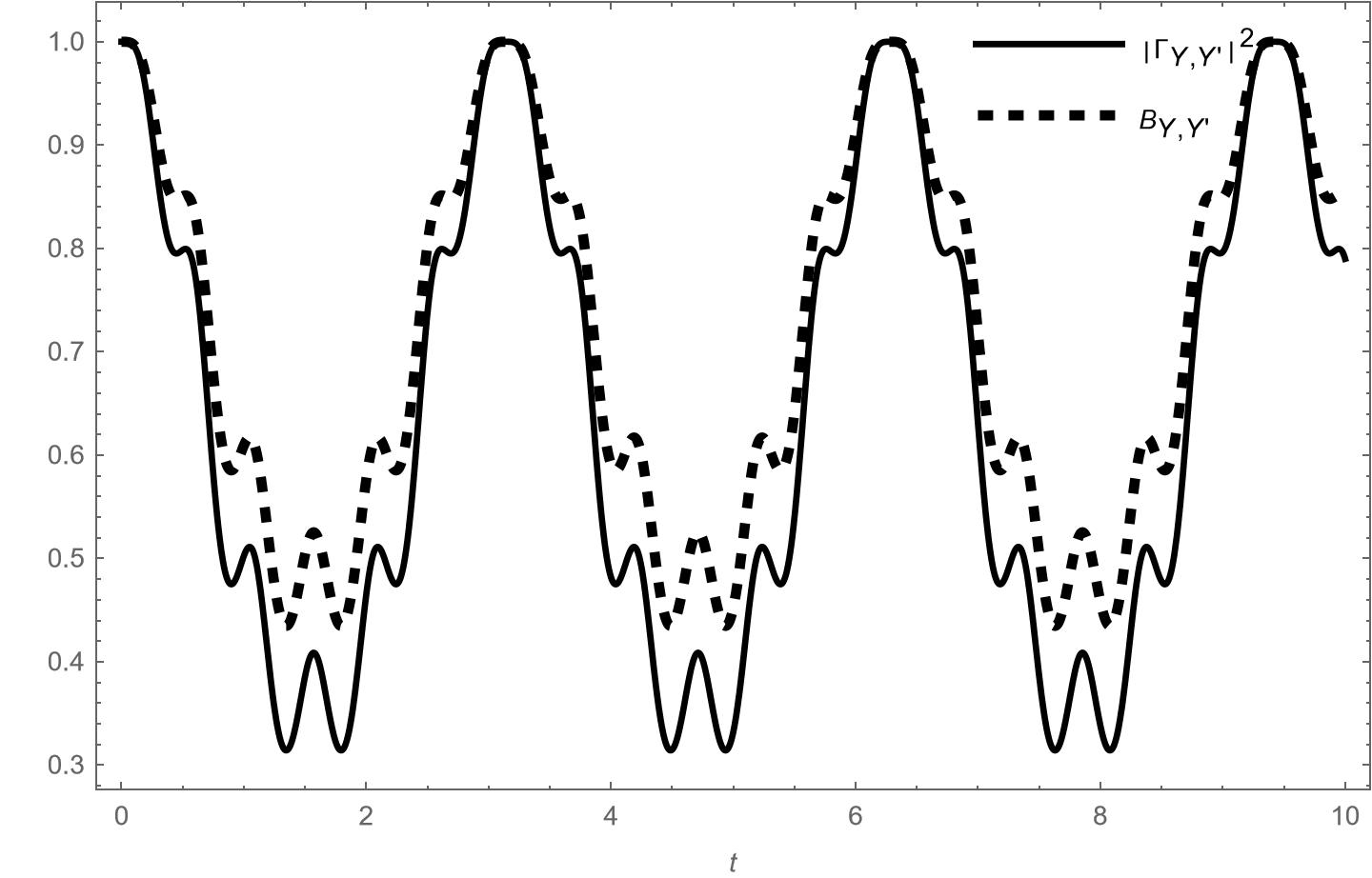}
  \caption{$\omega=4$ and $\Omega=6$. $t^e_{\text{min}}=\pi$.}
\end{subfigure}
\caption{\label{fig:Omega3omega2,Omega6omega4} Comparison between the objectivity markers having the same fractional relation between $\omega$ and $\Omega$ but with different frequency values, $(\omega,\Omega)=(4,6)$ and $(\omega,\Omega)=(2,3)$.}
\label{fig:same ratio}
\end{figure}
%%%%%%%%%%%%%%%%%%%%%%%%%%%%%%%%%%%%%%%%%%%%%%%%%%%%%%%

\begin{figure}[t!]
\includegraphics[width=1.0\linewidth, height=4cm]{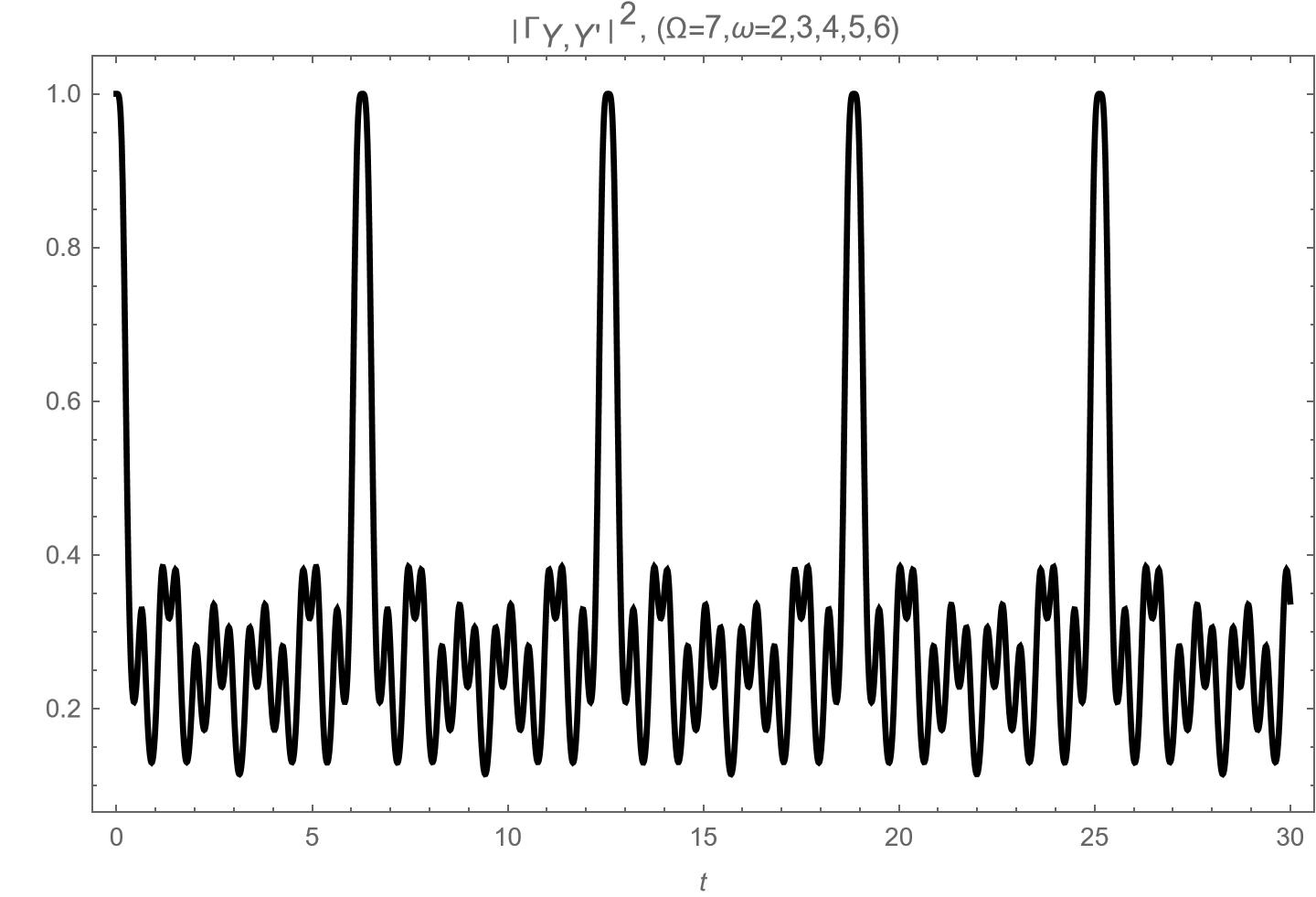} 
\caption{\label{fig:Total_Gamma_Fractional} The decoherence factor for the environment of the different fractional frequency relations with $\Omega=7$ and $\omega=2,3,4,5,6$.}
\end{figure}

\begin{figure}[t!]
\includegraphics[width=0.9\linewidth, height=4cm]{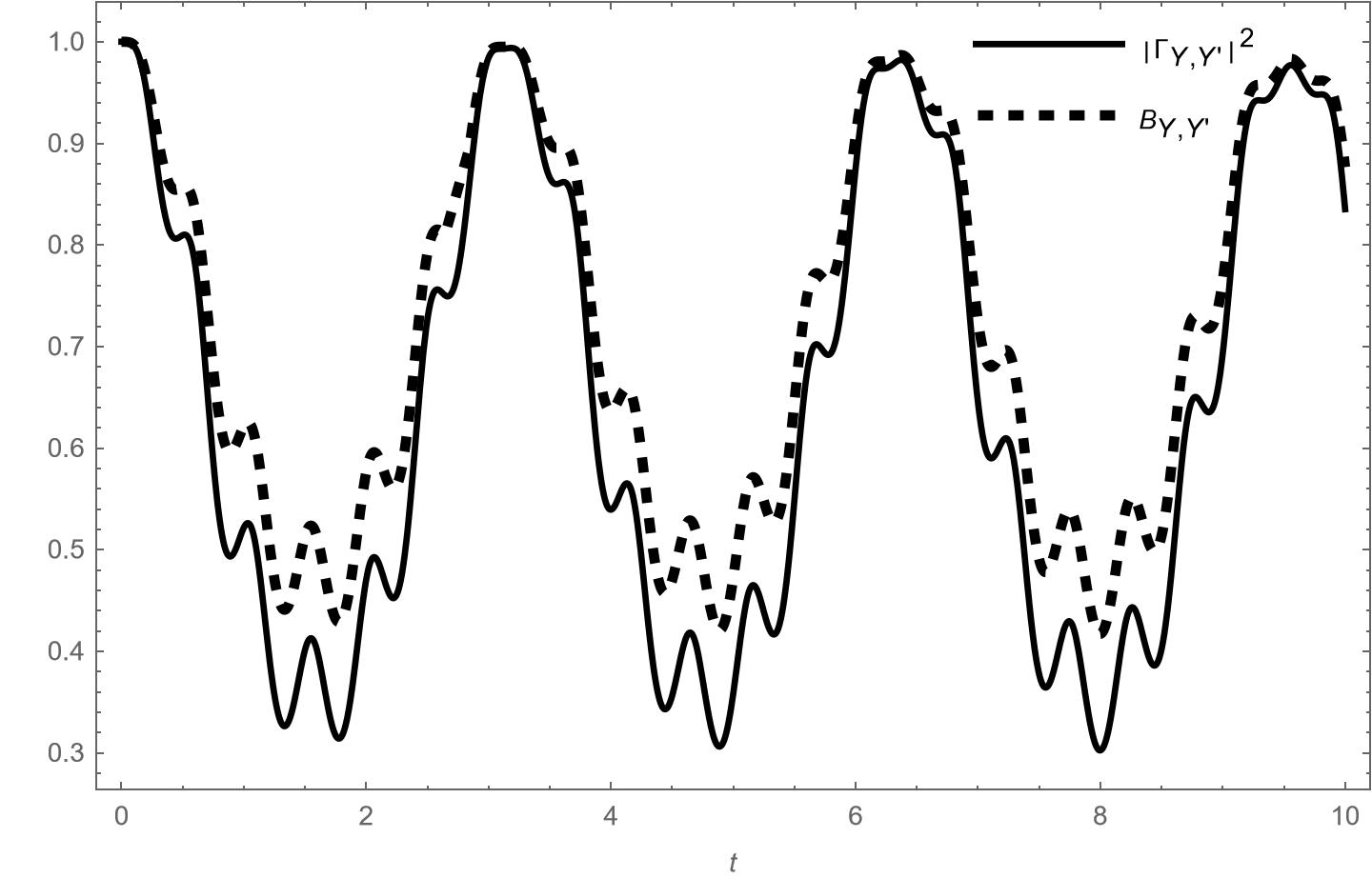} 
\caption{\label{fig:Nonfractional7.5} The objectivity markers for the environment having a non-fractional relation $\omega:\Omega=\sqrt{7^2+1}:\sqrt{5^2+1}$ to a central oscillator but close to a fractional relation $\omega:\Omega=5:7$ show the approximate non-objectivity in a short timescale.}
\end{figure}

\begin{figure}[t!]
\includegraphics[width=0.9\linewidth, height=4cm]{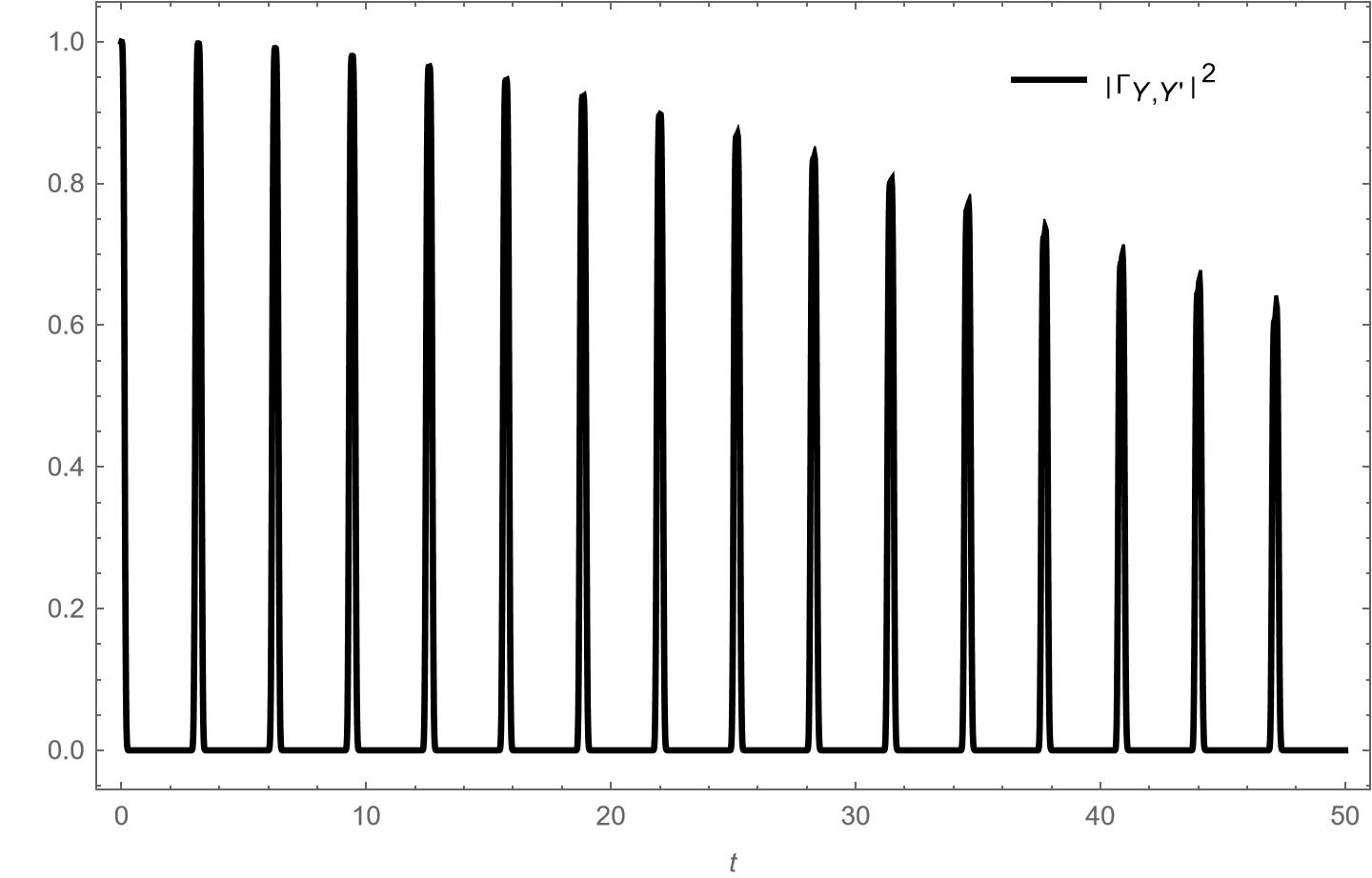} 
\caption{\label{fig:Nfcollective7.5} The decoherence factor for the environment consisting of oscillators with frequencies $\omega=\sqrt{25 + 0.01}, \sqrt{25 + 0.02},\sqrt{25 + 0.03}, \sqrt{25 + 0.04},\sqrt{25 + 0.05}$ with $\Omega=7$ in a timescale $t_s=50$.}
\end{figure}

\begin{figure}[t!]
\includegraphics[width=0.9\linewidth, height=4cm]{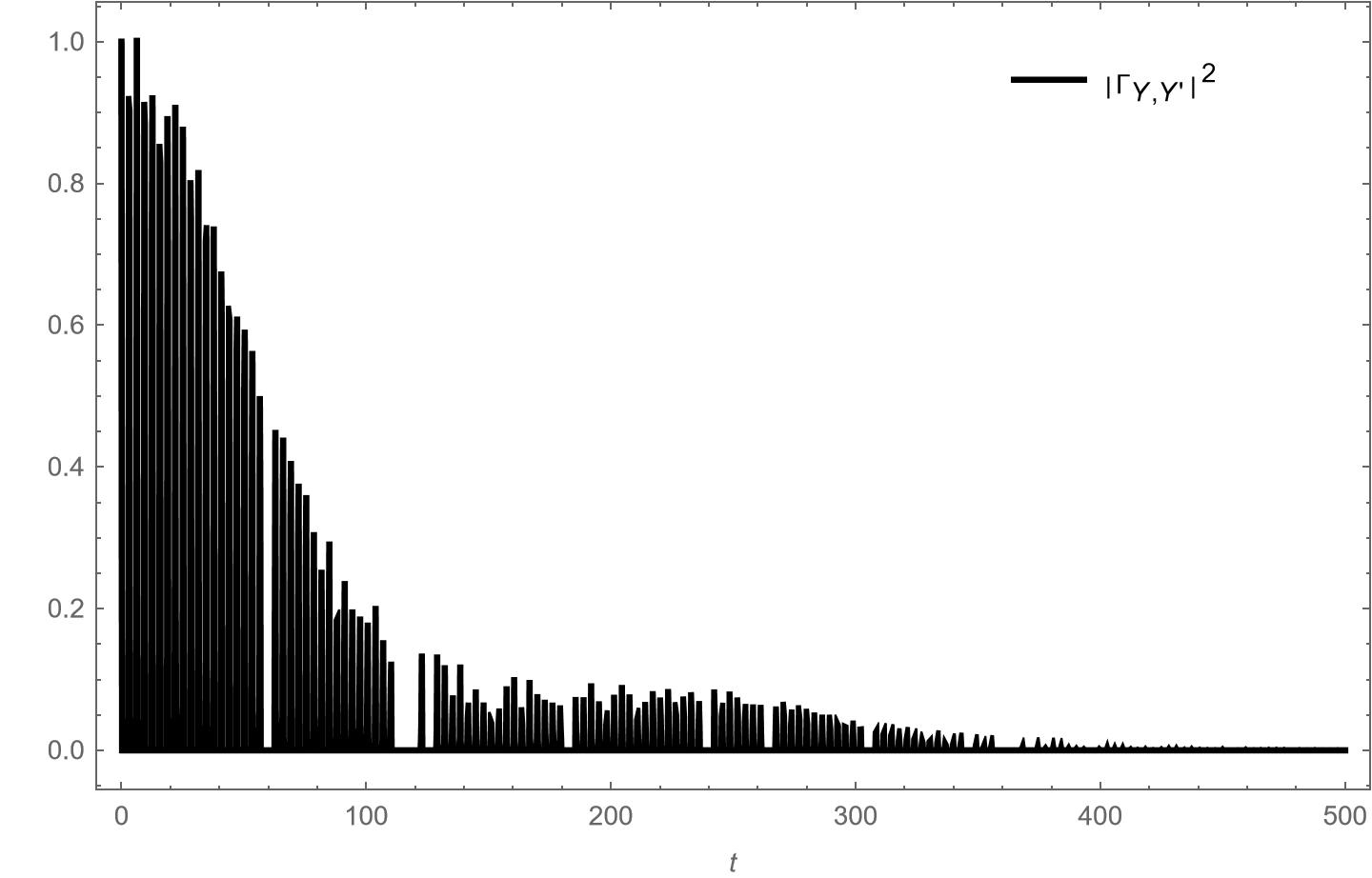} 
\caption{\label{fig:Nfcollective7.5largertime}The long time behavior for the decoherence factor for the environment consisting of oscillators with frequencies $\omega=\sqrt{25 + 0.01}, \sqrt{25 + 0.02},\sqrt{25 + 0.03}, \sqrt{25 + 0.04},\sqrt{25 + 0.05}$ with $\Omega=7$ in a timescale $t_s=500$.}
\end{figure}

In summary, for each given reduced fraction $f_{\text{re}}=\frac{m}{n}$, where positive integers $n,m\in\mathbb{N}$, there are the environmental frequency sets for the non-objectivity, when $n$ or $m$ is even, $\mathcal{W}^e_n=\{\omega_{n,m}:\omega_{n,m}=\frac{m}{n}\Omega,m\in\mathbb{N},m>0\}$ which provides the same series of time points for the non-objectivity, $\mathcal{T}^e_n=\{t^{n,m}_p:t^{n,m}_p=\frac{2n\pi}{\Omega}p,p\in\mathbb{N}\}$ for non-objectivity points. If $n$ and $m$ are both odd number, $\mathcal{W}^o_n=\{\omega_{n,m}:\omega_{n,m}=\frac{m}{n}\Omega,m\in\mathbb{N},m>0\}$ and $\mathcal{T}^o_n=\{t^{n,m}_p:t^{n,m}_p=\frac{n\pi}{\Omega}p,p\in\mathbb{N}\}$ for the non-objectivity points. If the environment is any mixture of $\mathcal{W}^e_n$ and $\mathcal{W}^o_l$, there should exist overlap between them and hence there is no objectivity. Fig. \ref{fig:Total_Gamma_Fractional} shows that if the environmental systems belong to $\mathcal{W}^e_n$ or $\mathcal{W}^o_n$, the objectivity cannot occur.
%%%%%%%%%%%%%%%%%%%%%%%%%%%%%%%%%%%%%%%%%%%%%%%%%%%%%%%
%%%%%%%%%%%%%%%%%%%%%%%%%%%%%%%%%%%%%%%%%%%%%%%%%%%%%%%%%
The periodicity of the objectivity markers in the fractional relation form a typical timescale of the system. Even if the environment does not follow an exact fractional relation with $\Omega$, it can form an approximate non-objectivity. Figs. \ref{fig:Nonfractional7.5} and \ref{fig:Nfcollective7.5} show the approximate periodicity of a decoherence factor from the non-fractional relation but Fig. \ref{fig:Nfcollective7.5largertime} shows that for a large timescale it has the objectivity. One way to achieve the objectivity is to give the environment a non-fractional frequency relation with a central oscillator. On the other hand, one can use a beating effect to use the environmental frequency close to the central one. If $|\Omega-\omega_k|=\Delta\omega_k\ll1$, $\frac{2\pi}{\Delta\omega_k}$ can be a typical timescale.

%%%%%%%%%%%%%%%%%%%%%%%%%%%%%%%%%%%%%%%%%%%%%%%%%%%%%%%
%%%%%%%%%%%%%%%%%%%%%%%%%%%%%%%%%%%%%%%%%%%%%%%%%%%%%%%% 

%%%%%%%%%%%%%%%%%%%%%%%%%%%%%%%%%%%%%%%%%%%%%%%%%%%%%%%%%%

%%%%%%%%%%%%%%%%%%%%%%%%%%%%%%%%%%%%%%%%%%%%%%%%%%%%%%%%%%
%%%%%%%%%%%%%%%%%%%%%%%%%%%%%%%%%%%%%%%%%%%%%%%%%%%%%%%

%%%%%%%%%%%%%%%%%%%%%%%%%%%%%%%%%%%%%%%%%%%%%%%%%%%%%%%
\subsection{Classical trajectory}\label{Classical trajectory}
\begin{figure}[t!]
\includegraphics[width=0.9\linewidth, height=4cm]{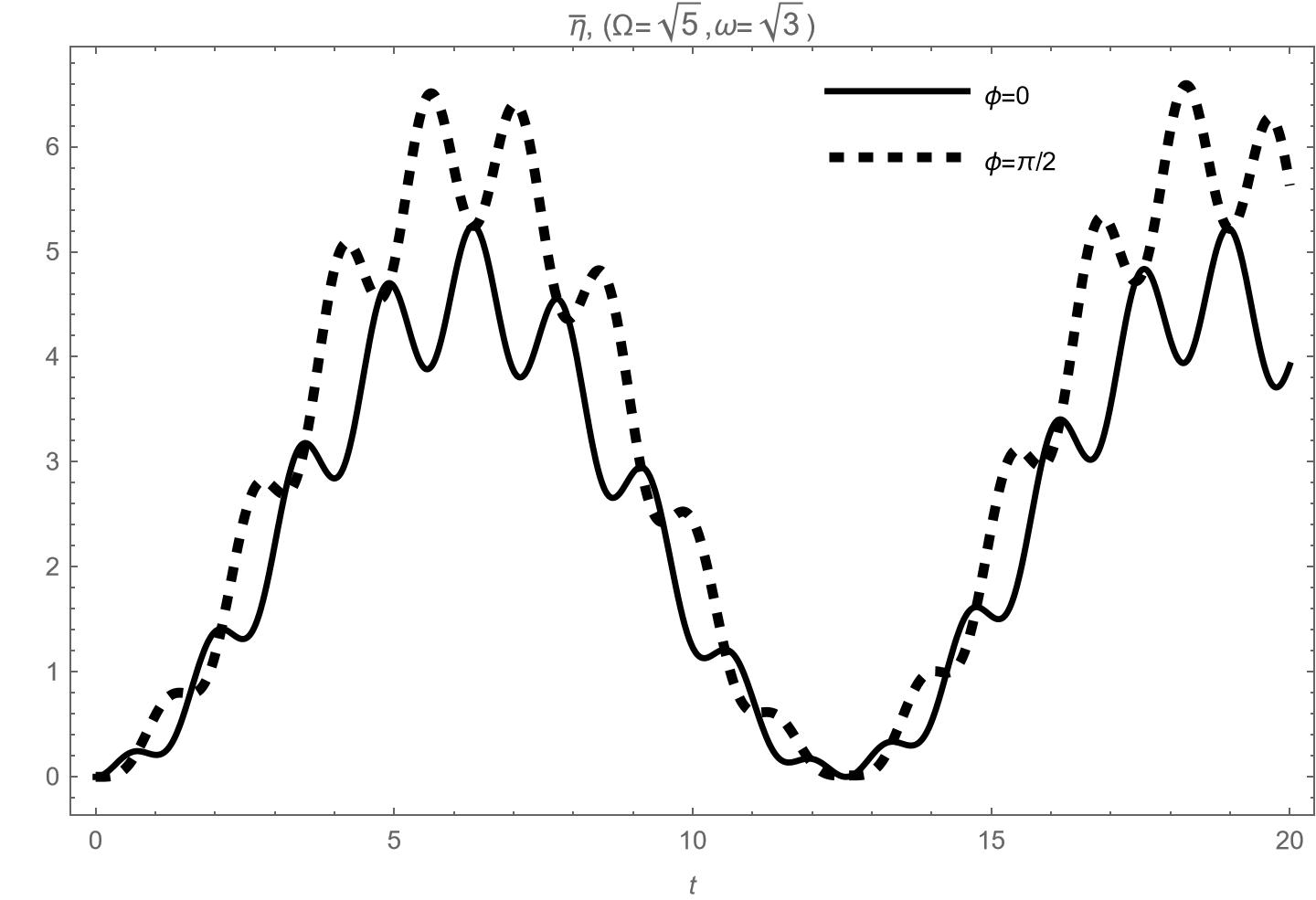}
\caption{\label{fig:phase in objectivity Omega>omega} For $\Omega>\omega$ ($\omega=\sqrt{3},~\Omega=\sqrt{5}$), $\bar{\eta}$ is maximized at $\phi=\frac{\pi}{2}$ and minimized at $\phi=0$.}
\includegraphics[width=0.9\linewidth, height=4cm]{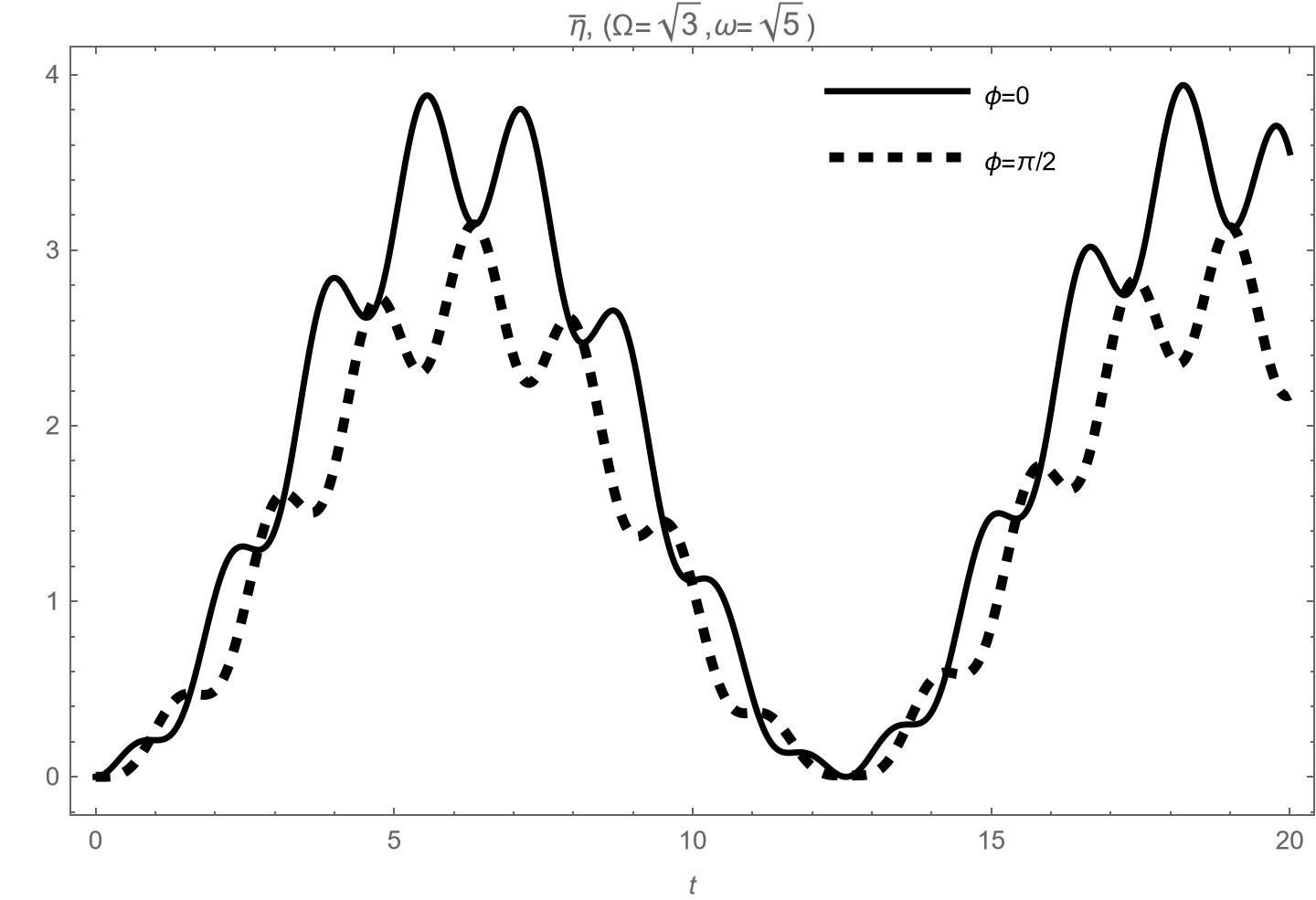}
  \caption{\label{fig:phase in objectivity Omega<omega}For $\Omega<\omega$ ($\omega=\sqrt{5},~\Omega=\sqrt{3}$), $\bar{\eta}$ is maximized at $\phi=0$ and minimized at $\phi=\frac{\pi}{2}$.}
\end{figure}

The fractional frequency conditions Eq. \eqref{even condition} and Eq. \eqref{odd condition} are independent of a phase in a classical trajectory. But the phase affects the objectivity in addition to any other objectivity conditions. It has not been completely clear in the boson-spin model why objectivity improves as the phase $\phi$ $(0\leq\phi\leq\pi/2)$ in a classical trajectory of a central oscillator increases \cite{Lee:2024iso,Lee:2024idx}. Here for the QBM model we show this explicitly. From Eq. \eqref{eta}, it is worth recognizing that the relevant quantity $\bar{\eta}_k\equiv\eta_k/g_kQ_k$ is a sum of two complex numbers $-e^{-i\omega t}v_k$ and $v_k(0)$:
\begin{align}\label{eta:v}
    \bar{\eta}_k=-e^{-i\omega t}v_k+v_k(0),
\end{align}
where
\begin{align}\label{v}\nonumber
    v_k&\equiv \cos(\Omega t+\phi)+i\frac{\Omega}{\omega_k}\sin(\Omega t+\phi),\\
v_k(0)&\equiv\cos\phi+i\frac{\Omega}{\omega_k}\sin\phi
\end{align}
and the magnitude of $\bar{\eta}_k$ is bounded by
\begin{align}\label{eta bar bound}\nonumber
    0\leq|\bar{\eta}_k|^2\leq\frac{4\Omega^2}{\omega^2_k},~(\Omega>\omega),\\
      0\leq|\bar{\eta}_k|^2\leq 1,~(\Omega<\omega).
\end{align}
The lower bound of $0$ occurs at the fractional relation between $\omega_k$ and $\Omega$. When $-e^{-i\omega t}v_k$ is anti-parallel to $v_k(0)$ and their magnitudes are maximized in Eq. \eqref{eta:v}, we find the upper bound at
\begin{align}\label{|v0|}
    |v_k|=|v_k(0)|=\sqrt{\left(\frac{\Omega^2}{\omega^2_k}-1\right)\sin^2\phi+1}.
\end{align}
This means that for the region $0\leq\phi\leq\pi/2$, as $\phi$ increases, $|\eta_k|^2$ increases for $\Omega>\omega_k$, while $\phi$ increases, $|\eta_k|^2$ decreases for $\Omega<\omega_k$. The conditions to maximize $|\bar{\eta}_k|^2$ are
\begin{align}\label{phi maximum eta}\nonumber
   |\bar{\eta}_k|^2_{\text{max}}&=\frac{\Omega^2}{\omega^2_k}~\text{at}~\phi=\frac{\pi}{2},~ (\Omega>\omega_k),\\
    |\bar{\eta}_k|^2_{\text{max}}&=1~\text{at}~\phi=0,~(\Omega<\omega_k).
\end{align}
Eq. \eqref{phi maximum eta} shows that the phase dependence of objectivity is determined by the relative energy scale between the central and environmental oscillators. Figs. \ref{fig:phase in objectivity Omega>omega} and \ref{fig:phase in objectivity Omega<omega} illustrate that $\bar{\eta}$ is maximized at $\phi=\pi/2$ and minimized at $\phi=0$ for $\Omega>\omega_k$ whereas it is minimized at $\phi=\pi/2$ and maximized at $\phi=0$ for $\Omega<\omega_k$.

%%%%%%%%%%%%%%%%%%%%%%%%%%%%%%%%%%%%%%%%%%%%%%%%%%%%%%%%%%
%%%%%%%%%%%%%%%%%%%%%%%%%%%%%%%%%%%%%%%%%%%%%%%%%%%%%%%
%%%%%%%%%%%%%%%%%%%%%%%%%%%%%%%%%%%%%%%%%%%%%%%%%%%%%%%%%%
%%%%%%%%%%%%%%%%%%%%%%%%%%%%%%%%%%%%%%%%%%%%%%%%%%%%%%%%%%
\subsection{Objectivity timescales}
\begin{figure}[t!]
\includegraphics[width=0.9\linewidth, height=4cm]{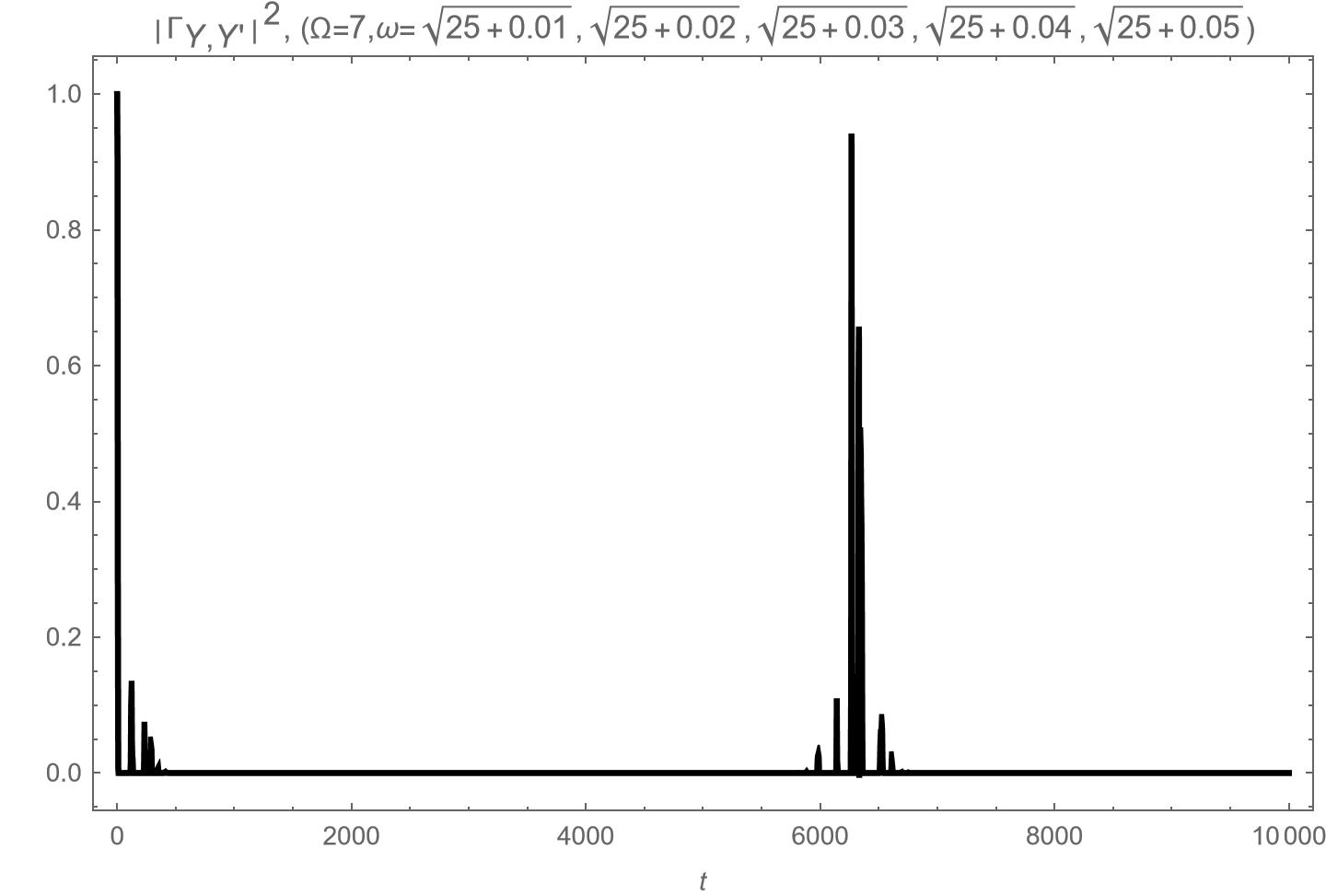}
\caption{\label{fig:Nfcollective7.5_ts10000} The decoherence factor for the environment consisting of oscillators with frequencies $\omega=\sqrt{25 + 0.01}, \sqrt{25 + 0.02},\sqrt{25 + 0.03}, \sqrt{25 + 0.04},\sqrt{25 + 0.05}$ with $\Omega=7$ in a timescale $t_s=10000$.}
\end{figure}

 The earliest non-objectivity time in Eq. \eqref{earliest even time} for a system under the fractional frequency relations, Eq. \eqref{even condition} and Eq. \eqref{odd condition}, can be regarded as a typical objectivity timescale. Within such a timescale the objectivity relies on the given physical parameters. For QBM under a recoilless limit, we numerically showed that it does not seem to be possible to obtain the decay for objectivity makers without any recurrence as $t\to\infty$ with the finite number of oscillators in the environment. When the objectivity markers are periodic, they do not decay at all. However, even if they are non-periodic due to non-fractional frequency relation between $\omega_k$ and $\Omega$, the approximate periodicity exists in finite time and objectivity markers cannot asymptotically decay. This implies that the objectivity can be only meaningful in a given timescale but not in an infinite timescale. Fig. \ref{fig:Nfcollective7.5_ts10000} gives the example that the non-objectivity reappears for a longer timescale, though in a shorter timescale it shows a good objectivity. In a timescale $t_s=10000$ the decoherence factor in the non-fractional relation has the recurrent amplitude, which does not appear in the timescale $t_s=500$ shown in Fig. \ref{fig:Nfcollective7.5largertime}. 
 %%%%%%%%%%%%%%%%%%%%%%%%%%%%%%%%%%%%%%%%%%%%%%%%%%%%%%%
%%%%%%%%%%%%%%%%%%%%%%%%%%%%%%%%%%%%%%%%%%%%%%%%%%%%%%%
\section{Conclusion}
This study is motivated by re-examining the previous objectivity studies in the QBM model under the recoilless limit with an initial thermal state \cite{Tuziemski:2015,Paz:2009,Lee:2023ozm}. We found two main results. One is the existence of objectivity timescales. The other is how classical trajectory of a central system affects objectivity. 

The objectivity timescales as a sign of non-objectivity in QBM under the recoilless limit exist with the finite number of the environmental oscillators. They have several different sources, independent of a classical trajectory of a central oscillator. One is a beating effect, when a central frequency is close to those from the environment, whose frequency is difference between them. Another is when a central frequency has a fractional relation with those of the environment. The resultant period of objectivity markers is some multiple of a central or environmental period. However, the absence of these non-objectivity sources does not give objectivity. However, we found that with the finite number of oscillators there exists a resurrection of non-objectivity independent of those origins, which requires further deeper study. 

On the other hand, we re-derived the objectivity markers including a general phase $\phi$ in the classical trajectory of a central oscillator as an exponential decay form, which is the same as those obtained in \cite{Tuziemski:2015, Lee:2023ozm}. We showed why an increased phase in a classical trajectory of a central oscillator leads to increased objectivity. The boson-spin model \cite{Lee:2024iso} has the same phase relation with objectivity. For QBM this phase relation works when a central frequency is greater than the environmental ones and for the opposite case the phase increases the objectivity decreases.

We conclude that objectivity with the finite numbers of environments is not defined only on a given timescale, whether the environment is in the fractional relation or not. 

We expect that on quantum Darwinism our non-objectivity sources with the finite number of environmental oscillators will lead a redundancy fall of the encoded information in the environment around objectivity timescale. It is worthwhile to confirm this anticipation in the future research. 

Nonetheless, our conclusion does not imply the failure of quantum Darwinism. As we know that in principle, any system interacts with the rest of the Universe and the non-objectivity recurrent time interval is effectively infinite. Moreover, we expect that such recurrent non-objectivity is not only for QBM but for any open quantum system with a finite number of environmental systems, rather than the defect of quantum Darwinism. This claim needs to be proved with a theoretical rigor.

 In the future, for the further completeness of the objectivity analysis for QBM, it is worth considering the corrections to the recoilless limit. On the other hand, to investigate objectivity for the mutually interacting environment would be an important but challenging work. Apart from the mutual interactions, to study how initially correlated environments affect objectivity would be an interesting and feasible research topic. For initially correlated environments the multiplicativity property for objectivity markers cannot be applied and it is necessary to compute the objectivity markers for the entire environment as a whole. However, for this case, the non-objectivity condition due to periodicity from the frequency relations in objectivity markers we found here will not be an issue. I anticipate that such a non-separable property would lead to better objectivity than separable cases.

\onecolumngrid\
\appendix
\label{appendix}
In this Appendix we provide the detailed calculations for the derivations for the objectivity markers with the help of the techniques used in Refs. \cite{Tuziemski:2015, Lee:2023ozm} but using the Floquet decomposition. In addition, in the last section \ref{appendix C}, we derive the time points where the objectivity markers are maximized. Note that in Appendix we drop the environmental index ``$k$'' since it is sufficient for the calculation to consider only a single environment.
%%%%%%%%%%%%%%%%%%%%%%%%%%%%%%%%%%%%%%%%%%%%%%%%%%%%%%%%%%%%%%%
\section{Decoherence factor}\label{appendix A}
In this section we derive a decoherence factor for QBM under the recoilless limit discussed in Sec.\ref{Recoilless limit} with a thermal initial state. 

It is noticeable that the Hamiltonian $H_{\text{eff}}$ for the environment given in Eq. \eqref{Heff:Y} cannot be simultaneously diagonalized with a thermal initial state, which commutes with the Hamiltonian for a harmonic oscillator. However, using the property that the unitary evolution operator $U_{\text{eff}}$ in Eq. \eqref{Ueff} is expressed as a product of the displacement operators and the unitary evolution operator for a harmonic oscillator and the diagonal representation in the coherent basis defined below Eq. \eqref{A:diagonal representation}, a decoherence factor can be exactly calculable. It is known that any density matrix of a one dimensional harmonic oscillator can be written in a diagonal form in a coherent state basis \cite{Sudarshan:1963ts,Mehta:1967}. According to this representation, our initial density matrix $\rho_0$ can be expressed in a diagonal form in the coherent basis $\{|v\rangle\}$:
\begin{align}\label{A:diagonal representation}
    \rho_0=\int d^2v \varphi(v)|v\rangle\langle v|.
\end{align} 
In general, for a given density matrix, the coefficient $\varphi(v)$ in Eq. \eqref{A:diagonal representation} is determined. For a thermal state for a harmonic oscillator, $\rho_{\text{th}}=e^{-\beta H}/\text{Tr}[e^{-\beta H}]$, where $\beta=1/k_BT$, $\varphi_{\text{th}}(v)$ is known as \cite{Mehta:1967}
\begin{align}\label{A:phi thermal}
   \varphi_{\text{th}}(v)=\frac{\mu_\lambda}{\pi}e^{-\mu_\lambda|v|^2},
\end{align}
where $\mu_\lambda\equiv e^\lambda-1$ and $\lambda\equiv \omega/k_BT$.
On the other hand, a unitary evolution operator for QBM in the Floquet decomposition, $U_{\text{eff}}\equiv U_Y$ in Eq. \eqref{Ueff}, is written in terms of the displacement operators:
\begin{align}\label{A:Uy}\nonumber
    U_Y&=e^{-iK_Y}e^{-iH_Ft}e^{iK_{Y0}}\\
    &=e^{-iK_Y}\left[e^{-iH_Ft}e^{iK_{Y0}}e^{iH_Ft}\right]e^{-iH_Ft}\\
    &=D[-Y\alpha]D[Ye^{i\omega t}\alpha_0]e^{-iH_Ft}.\nonumber
\end{align}
Here $K_Y$ has a periodicity due to the angular frequency $\Omega$ in $H_{\text{eff}}$  in Eq. \eqref{Heff:Y} and is given in Eq. \eqref{KY} as 
\begin{align}
    -iK_Y=-Y(\alpha a^\dagger-\alpha^*a),
\end{align}
where the displacement operator $D(\alpha)\equiv e^{\alpha a^\dagger-\alpha^*a}$
and $e^{-iH_Ft}e^{iK_{Y0}}e^{iH_Ft}$ is the Heisenberg picture of $e^{iK_{Y0}}$ with a negative time. 
 Using the property $D(\alpha)D(\beta)=e^{(\alpha\beta^*-\alpha^*\beta)/2}D(\alpha+\beta)$, Eq. \eqref{A:Uy} leads $U_Y|v\rangle$ to be another coherent state up to some phase $e^{\theta_Y}$ as
\begin{align}\label{A:unitary evolution on a coherent state}\nonumber
    U_Y|v\rangle
   &=D[-Y\alpha]D[Ye^{i\omega t}\alpha_0]D(e^{i\omega t}v)|0\rangle\\\nonumber
   &=e^{-Y^2(e^{-i\omega t}\alpha\alpha^*_0-e^{i\omega t}\alpha^*\alpha_0)/2}D[Ye^{i\omega t}\eta]D[e^{i\omega t}v]\\\nonumber
    &=e^{-Y^2(e^{-i\omega t}\alpha\alpha^*_0-e^{i\omega t}\alpha^*\alpha_0)/2}e^{Y(\eta v^*-\eta*v)/2}D(\gamma_Y)\\
    &=e^{\theta_Y}D(\gamma_Y),
\end{align}
where $\theta_Y$ is 
\begin{align}\label{A:theta_Y}
    \theta_Y
    &=e^{-Y^2(e^{-i\omega t}\alpha\alpha^*_0-e^{i\omega t}\alpha^*\alpha_0)/2}e^{Y(\eta v^*-\eta*v)/2},
\end{align}
\begin{align}\label{A: gamma_Y}
    \gamma_Y\equiv e^{i\omega t}(Y\eta+v)
\end{align}
and $\eta=-e^{-i\omega t}\alpha+\alpha_0$ from Eq. \eqref{eta}. Only the $v$ dependence in $\theta_Y$ in Eq. \eqref{A:theta_Y}, $\theta_Y(v,\alpha)$,
\begin{align}\label{A:theta relevant}\nonumber
    \theta_Y(v,\alpha)&=\theta_Y-\theta_Y(\alpha)\\
    &=\frac{Y}{2}(\eta v^*-\eta^*v )
\end{align}
is relevant to a decoherence factor and the rest $\theta_Y(\alpha)$
\begin{align}\label{A:theta irrelevant}
    \theta_Y(\alpha)=-\frac{Y^2}{2}(e^{-i\omega t}\alpha\alpha^*_0-e^{i\omega t}\alpha^*\alpha_0)
\end{align}
is irrelevant for our calculation.
Using Eq. \eqref{A:theta relevant} and Eq. \eqref{A:theta irrelevant}, we write the operator $A\equiv U_Y\rho_0Y^\dagger_{Y'}$ defining a decoherence factor as $|\Gamma_{Y,Y'}|^2=|\text{Tr}[A]|^2$:
\begin{align}\label{A:A}
    A=e^{\theta_Y(\alpha)+\theta^*_{Y'}(\alpha)}\int d^2 v \varphi(v)e^{\theta_Y(v,\alpha)+\theta^*_{Y'}(v,\alpha)}|\gamma_Y\rangle\langle \gamma_{Y'}|,
\end{align}
where
\begin{align}\label{A:theta relevant total}
  \theta_Y(v,\alpha)+ \theta^*_{Y'}(v,\alpha)&=\frac{1}{2}(Y-Y')(\eta v^*-\eta^*v ).
\end{align}
In Eq. \eqref{A:A}, noticing the trace $\text{Tr}(|\gamma_Y\rangle\langle \gamma_{Y'}|)=\langle\gamma_{Y'}|\gamma_Y\rangle$ and a well-known expression for the coherent state overlap, $\langle\alpha|\beta\rangle=e^{-\frac{1}{2}(|\alpha|^2+|\beta|^2-2\alpha^*\beta)}$, $\langle\gamma_{Y'}|\gamma_Y\rangle$ is expressed with Eq. \eqref{A: gamma_Y},
\begin{align}\label{A:coherent state overlap}\nonumber
    \langle\gamma_{Y'}|\gamma_Y\rangle&=e^{-\frac{1}{2}(|\gamma_{Y}|^2+|\gamma_{Y'}|^2-2\gamma^*_{Y'}\gamma_Y)}\\\nonumber
    &=e^{-\frac{1}{2}[|Y\eta+v|^2+|Y'\eta+v|^2-2(Y'\eta^*+v^*)(Y\eta+v)]}\\\nonumber
    &=e^{-\frac{1}{2}[(Y-Y')^2|\eta|^2+Y\eta v^*+Y\eta^*v+Y'\eta v^*+Y'\eta^*v-2(Y' \eta^* v+Y\eta v^*))]}\\\nonumber
    &=e^{-\frac{1}{2}[(Y-Y')^2|\eta|^2-Y\eta v^*+Y\eta^*v+Y'\eta v^*-Y'\eta^*v]}\\
    &=e^{-\frac{1}{2}(Y-Y')^2|\eta|^2}e^{\frac{1}{2}(Y-Y')(\eta v^*-\eta^*v)}.
\end{align}
From Eq. \eqref{A:A}, using Eq. \eqref{A:theta relevant total}, $\text{Tr}[A]$ is
\begin{align}\label{A:trace A}
   \text{Tr}[A]&=e^{\theta_Y(\alpha)+\theta^*_{Y'}(\alpha)}e^{-\frac{1}{2}(Y-Y')^2|\eta|^2}\int d^2 v\varphi(v)e^{\Delta Y(\eta v^*-\eta^*v)},
\end{align}
where $\Delta Y\equiv Y-Y'$.
 In $\text{Tr}[A_{\text{th}}]$ from Eq. \eqref{A:trace A}, the integration relevant part over $v\equiv x+iy$ with $d^2v\equiv dxdy$ and $\eta\equiv\eta_r+i\eta_i$ is
\begin{align}\label{A:Ath integral}\nonumber
    \int d^2v\varphi_{\text{th}}(v)e^{\Delta Y(\eta v^*-\eta^* v)}
   &=\frac{\mu_\lambda}{\pi}\int dxdy e^{-\mu_\lambda x^2+\Delta Y(\eta-\eta^*)x}e^{-\xi y^2-i\Delta Y(\eta+\eta^*)y}\\\nonumber
   &=\frac{\mu_\lambda}{\pi}\int dxdy e^{-\mu_\lambda x^2+2i\Delta Y\eta_ix}e^{-\mu_\lambda y^2-2i\Delta Y\eta_r y}\\\nonumber
   &=\frac{\mu_\lambda}{\pi}\int dxdy e^{-\mu_\lambda (x-i\Delta Y \eta_i/\zeta)^2-\frac{(\Delta Y)^2\eta^2_i}{\mu_\lambda}}e^{-\mu_\lambda (y+i\Delta Y \eta_r\mu_\lambda)^2-\frac{(\Delta Y)^2\eta^2_r}{\mu_\lambda}}\\
   &=e^{-\frac{(\Delta Y)^2}{\mu_\lambda}|\eta|^2}.
\end{align}
Plugging Eq. \eqref{A:Ath integral} into Eq. \eqref{A:trace A} and using $\mu_\lambda\equiv e^\lambda-1$, the decoherence factor $|\Gamma_{Y,Y'}|^2$ is 
\begin{align}\nonumber
    |\Gamma_{Y,Y'}|^2&=|\text{Tr}[A_{\text{th}}]|^2\\\nonumber
    &=e^{-(\Delta Y)^2|\eta|^2}e^{-\frac{(\Delta Y)^2}{\mu_\lambda}|\eta|^2}\\\nonumber
    &=e^{-(1+2\mu_\lambda^{-1})(\Delta Y)^2|\eta|^2}\\
     &=e^{-\frac{e^\lambda+1}{e^\lambda-1}(\Delta Y)^2|\eta|^2}\\\nonumber
     &=e^{-\coth[\lambda/2](\Delta Y)^2|\eta|^2}.\nonumber
\end{align}
Using $\lambda\equiv \omega/k_BT$, a decoherence factor with a thermal state is explicitly expressed
\begin{align}\label{A:decoherence factor}
    |\Gamma_{Y,Y'}|^2
     &=e^{-\coth\left[\frac{\omega}{2k_BT}\right](\Delta Y)^2|\eta|^2}.
\end{align}
This is the same exponential form as in \cite{Tuziemski:2015}, containing a general phase $\phi$ in $\eta$.
%%%%%%%%%%%%%%%%%%%%%%%%%%%%%%%%%%%%%%%%%%%%%%%%%%%%%%%%%%%%%%%%%%
\section{Generalized overlap}\label{appendix B}
In this section, similarly to the previous section we derive a generalized overlap with a thermal state. Due to the structural difference from a decoherence factor, computing a generalized overlap is less straightforward than a decoherence factor. We introduce the operator $B$ below defining a generalized overlap $B_{Y,Y'}=|\text{Tr}\sqrt{B}|^2$ as
\begin{align}
    B&\equiv\sqrt{\rho_0}U^\dagger_Y U_{Y'}\rho_0U^\dagger_{Y'} U_{Y}\sqrt{\rho_0}.
\end{align}
With $K_Y$ in Eq. \eqref{KY}, $U_Y$ in Eq. \eqref{A:Uy} can be rewritten as
\begin{align}\label{A:Uy2}\nonumber
    U_Y&=e^{-iH_Ft}\left[e^{iH_Ft}e^{-iK_Y}e^{-iH_Ft}\right]e^{iK_{Y0}}\\
    &=e^{-iH_Ft}D[-Ye^{-i\omega t}\alpha]D[Y\alpha_0].
\end{align}
$U_{Y,Y'}\equiv U^\dagger_Y U_{Y'}$ can be expressed as
\begin{align}\label{A:UYY'}\nonumber
    U_{Y,Y'}&\equiv U^\dagger_Y U_{Y'}\\\nonumber
    &=D[-Y'\alpha_0]D[Y'e^{-i\omega t}\alpha]D[-Ye^{-i\omega t}\alpha]D[Y\alpha_0]\\
    &=e^{\theta_Y(\alpha)+\theta^*_{Y'}(\alpha)}D[-Y'\eta]D[Y\eta]\\\nonumber
    &=e^{\theta_Y(\alpha)+\theta^*_{Y'}(\alpha)}D[(Y-Y')\eta],\nonumber
\end{align}
where $\eta=-e^{-i\omega t}\alpha+\alpha_0$ in Eq. \eqref{eta} and $e^{\theta_Y(\alpha)+\theta^*_{Y'}(\alpha)}$ defined in Eq. \eqref{A:theta irrelevant} does not contribute to $B$.
With $\rho_0=\int d^2v\varphi(v)|v\rangle\langle v|$,
\begin{align}\label{A:B}\nonumber
    B&=\int d^2 v\varphi(v)\rho^{1/2}_0D[\Delta Y\eta]|v\rangle\langle v|D^\dagger[\Delta Y\eta]\rho^{1/2}_0\\
    &=\int d^2 v\varphi(v)\rho^{1/2}_0|v+\Delta Y\eta\rangle\langle v+\Delta Y\eta|\rho^{1/2}_0.
\end{align}
For a thermal state for a harmonic oscillator, $\rho_{\text{th}}(\lambda)= (1-e^{-\lambda}) e^{-\lambda a^\dagger a}$, where $\lambda\equiv\omega/k_BT$, if $\sqrt{\rho_{\text{th}}(\lambda)}$ is expanded in the energy eigenbasis $|n\rangle$,
 \begin{align}\label{A:square root thermal}
     \sqrt{\rho_{\text{th}}(\lambda)}&=(1-e^{-\lambda})^{1/2}\sum_ne^{-\frac{1}{2}\lambda n}|n\rangle\langle n|.
 \end{align}
 Thus, plugging this expression Eq. \eqref{A:square root thermal} and using a projection of a coherent state on the energy eigenbasis $|n\rangle$ in the expansion  $|\alpha\rangle=e^{-|\alpha|^2/2}\sum\limits_{n=0}\frac{\alpha^n}{\sqrt{n!}}|n\rangle$ into $B$ in Eq. \eqref{A:B}, 
 \begin{align}\label{A:square root thermal on energy basis}\nonumber
     \rho^{1/2}_{\text{th}}|v+\Delta Y\eta\rangle&=(1-e^{-\lambda})^{1/2}\sum_ne^{-\frac{1}{2}\lambda n}|n\rangle\langle n|v+\Delta Y\eta\rangle\\\nonumber
     &=(1-e^{-\lambda})^{1/2}e^{-\frac{1}{2}|v+\Delta Y\eta|^2}\sum_ne^{-\frac{1}{2}\lambda n}\frac{(v+\Delta Y\eta)^n}{\sqrt{n!}}|n\rangle\\\nonumber
     &=(1-e^{-\lambda})^{1/2}e^{-\frac{1}{2}|v+\Delta Y\eta|^2}\sum_n\frac{[e^{-\frac{1}{2}\lambda }(v+\Delta Y\eta)]^n}{\sqrt{n!}}|n\rangle\\
     &=\epsilon\mu_\lambda^{1/2} e^{-\frac{1}{2}\mu_\lambda|w|^2}\sum_ne^{-|w|^2/2}\frac{w^n}{\sqrt{n!}}|n\rangle,
\end{align}
where $\epsilon\equiv e^{-\frac{1}{2}\lambda}$, $\mu_\lambda\equiv e^\lambda-1$ and $w\equiv \epsilon(v+\Delta Y\eta)$.
Re-expressing Eq. \eqref{A:square root thermal on energy basis} in the coherent state basis with $|w\rangle=e^{-|w|^2/2}\sum\limits_{n=0}\frac{w^n}{\sqrt{n!}}|n\rangle$,
\begin{align}\label{A:square root thermal on a coherent state}
     \rho^{1/2}_{\text{th}}|v+\Delta Y\eta\rangle
      &=\epsilon\mu_\lambda e^{-\frac{1}{2}\mu_\lambda|w|^2}|w\rangle
\end{align}
and using Eq. \eqref{A:phi thermal} and $\epsilon^2d^2v=d^2w$, $B$ in Eq. \eqref{A:B} becomes
\begin{align}\label{A:B2}\nonumber
    B&=\frac{\epsilon^2\mu_\lambda^2}{\pi}\int d^2ve^{-\zeta|v|^2-\epsilon^2\mu_\lambda|v+\Delta Y\eta|^2}|w\rangle\langle w|\\
    &=\frac{\mu_\lambda^2}{\pi}\int d^2we^{-\epsilon^{-2}\mu_\lambda|w-\epsilon\Delta Y\eta|^2-\mu_\lambda|w|^2}|w\rangle\langle w|.
\end{align}
The exponent inside the integral in Eq. \eqref{A:B2} is replaced with a completed square form:
\begin{align}\label{A:square recombine}\nonumber
    -\epsilon^{-2}\mu_\lambda|w-\epsilon\Delta Y\eta|^2-\mu_\lambda|w|^2&=-\epsilon^{-2}\mu_\lambda|w|^2-\mu_\lambda|w|^2+\epsilon^{-1}\mu_\lambda \Delta Y\eta w^*+\epsilon^{-1}\mu_\lambda \Delta Y\eta^* w-\mu_\lambda(\Delta Y)^2|\eta|^2\\\nonumber
    &=-(\epsilon^{-2}+1)\mu_\lambda\left[|w|^2-\frac{\epsilon^{-1}}{(\epsilon^{-2}+1)} \Delta Y\eta w^*+-\frac{\epsilon^{-1}}{(\epsilon^{-2}+1)} \Delta Y\eta^* w\right]-\mu_\lambda(\Delta Y)^2|\eta|^2\\\nonumber
    &=-(\epsilon^{-2}+1)\mu_\lambda\left|w-\frac{\epsilon^{-1}}{(\epsilon^{-2}+1)} \Delta Y\eta \right|^2-\frac{\mu_\lambda}{\epsilon^{-2}+1}(\Delta Y)^2|\eta|^2\\\nonumber
    &=-(e^{2\lambda}-1)\left|w-\frac{\epsilon^{-1}}{(\epsilon^{-2}+1)} \Delta Y\eta \right|^2-\tanh(\lambda/2)(\Delta Y)^2|\eta|^2\\
    &=-\mu_{2\lambda}|w-\Delta|^2-\tanh(\lambda/2)(\Delta Y)^2\eta^2,
\end{align}
where $\mu_{2\lambda}\equiv e^{2\lambda}-1$ and $\Delta\equiv \frac{\epsilon^{-1}}{(\epsilon^{-2}+1)} \Delta Y\eta$. Replacing the exponent in Eq. \eqref{A:B2} with Eq. \eqref{A:square recombine}, $B$ becomes
\begin{align}\label{A:B3}\nonumber
    B&=\frac{\mu_\lambda^2}{\pi}\int d^2we^{-\mu_{2\lambda}|w-\Delta|^2-\tanh(\lambda/2)(\Delta Y)^2\eta^2}|w\rangle\langle w|\\\nonumber
    &=e^{-\tanh(\lambda/2)(\Delta Y)^2|\eta|^2}\frac{\mu_\lambda^2}{\pi}\int d^2\tilde{w}e^{-\mu_{2\lambda}|\tilde{w}|^2}|\tilde{w}+\Delta\rangle\langle \tilde{w}+\Delta|\\
     &=e^{-\tanh(\lambda/2)(\Delta Y)^2|\eta|^2}\frac{\mu_\lambda^2}{\pi}D(\Delta)\left[\int d^2\tilde{w}e^{-\mu_{2\lambda}|\tilde{w}|^2}|\tilde{w}\rangle\langle \tilde{w}|\right]D^\dagger(\Delta).
\end{align}
Recalling the diagonal representation for a thermal state Eq. \eqref{A:diagonal representation}, $\rho_{\text{th}}=\int d^2v\varphi_{\text{th}}(v)|v\rangle\langle v|$ with $\varphi_{\text{th}}(v)=\frac{\mu_\lambda}{\pi}e^{-\mu_\lambda|v|^2}$, $B$ in Eq. \eqref{A:B3} can be written as
\begin{align}\label{A:B3}\nonumber
    B
   &=e^{-\tanh(\lambda/2)(\Delta Y)^2|\eta|^2}\frac{\mu_\lambda^2}{\mu_{2\lambda}}\frac{\mu_{2\lambda}}{\pi}D(\Delta)\left[\int d^2\tilde{w}e^{-\mu_{2\lambda}|\tilde{w}|^2}|\tilde{w}\rangle\langle \tilde{w}|\right]D^\dagger(\Delta)\\
   &=e^{-\tanh(\lambda/2)(\Delta Y)^2|\eta|^2}\frac{\mu_\lambda^2}{\mu_{2\lambda}}D(\Delta)\rho_{\text{th}}(2\lambda)D^\dagger(\Delta).
\end{align}
As seen in Eq. \eqref{A:square root thermal}, $\rho_{\text{th}}(2\lambda)$ is related to $\rho^2_{\text{th}}(\lambda)$ as
\begin{align}\label{A:squared thermal}\nonumber
     \rho^2_{\text{th}}(\lambda)&=(1-e^{-\lambda})^2\sum_ne^{-2\lambda n}|n\rangle\langle n|\\\nonumber
     &=\frac{(1-e^{-\lambda})^2}{(1-e^{-2\lambda})}(1-e^{-2\lambda})\sum_ne^{-2\lambda n}|n\rangle\langle n|\\
     &=\frac{\mu^2_{\lambda}}{\mu_{2\lambda}}\rho_{\text{th}}(2\lambda).
 \end{align}
 $B$ is expressed with $\rho^2_{\text{th}}(\lambda)$,
 \begin{align}\label{A:B4}
    B
   &=e^{-\tanh(\lambda/2)(\Delta Y)^2|\eta|^2}D(\Delta)\rho^2_{\text{th}}(\lambda)D^\dagger(\Delta).
\end{align}
Taking a square root on $B$ in Eq. \eqref{A:B4}, $\sqrt{B}$ is obtained as
\begin{align}\label{A:B square root}
    \sqrt{B}&=D(\Delta)e^{-\frac{1}{2}\tanh(\lambda/2)(\Delta Y)^2|\eta|^2}\rho_{\text{th}}(\lambda)D^\dagger(\Delta).
\end{align}
Using $\text{Tr}\rho_{\text{th}}(\lambda)=1$, the generalized overlap $B_{Y,Y'}=|\text{Tr}\sqrt{B}|^2$ is
\begin{align}\label{A:generalized overlap}
    B_{Y,Y'}=e^{-\tanh(\frac{\omega}{2k_BT})(\Delta Y)^2|\eta|^2}.
\end{align}

%%%%%%%%%%%%%%%%%%%%%%%%%%%%%%%%%%%%%%%%%%%%%%%%%%%%%%%%%%%%%
\section{$\eta$}\label{appendix C}
The common quantity on the exponent in the objectivity markers, $\eta$ in Eq. \eqref{eta}, is an important factor in the objectivity for QBM.
\subsection{Zeros of $\eta$}
This section searches for a condition satisfying $\eta=0$, namely the non-objectivity condition. $\eta$ is the sum of two complex numbers:
\begin{align}\label{A:eta}
    \eta&=gQ\left[-e^{-i\omega t}\left(c+i\frac{\Omega}{\omega}s\right)+c_0+i\frac{\Omega}{\omega}s_0\right],
\end{align}
with $c\equiv\cos(\Omega t+\phi)$, $s\equiv \sin(\Omega t+\phi)$, $c_0\equiv\cos\phi$, $s_0=\sin\phi$ and
\begin{align}\label{A:Q}
    Q\equiv\sqrt{\frac{\omega}{2m(\omega^2-\Omega^2)^2}}.
\end{align}
Consider the relevant complex number $\bar{\eta}$ to find out zeros for $t$:
\begin{align}\label{A:eta bar}
    \bar{\eta}\equiv\frac{\eta}{gQ}=-e^{-i\omega t}\left(c+i\frac{\Omega}{\omega}s\right)+\left(c_0+i\frac{\Omega}{\omega}s_0\right).
\end{align}
We represent the complex numbers in Eq. \eqref{A:eta bar} in the Euler's form:
\begin{align}
    c+i\frac{\Omega}{\omega}s=Ae^{i\delta},~c_0+i\frac{\Omega}{\omega}s_0=A_0e^{i\delta_0},
\end{align}
where
\begin{align}\label{A:amplitudes angles}\nonumber
A=\sqrt{c^2+\frac{\Omega^2}{\omega^2}s^2},
&~A_0=\sqrt{c^2_0+\frac{\Omega^2}{\omega^2}s^2_0},\\
    \tan\delta=\frac{\omega}{\Omega}\tan(\Omega t+\phi),&~\tan\delta_0=\frac{\omega}{\Omega}\tan\phi.
\end{align}
$\bar{\eta}=0$ in Eq. \eqref{A:eta bar} is written as 
\begin{align}\label{A:eta zero}\nonumber
  \bar{\eta}&=-e^{-i\omega t}Ae^{i\delta}+A_0e^{i\delta_0}=0\\
  &\to \frac{A}{A_0}e^{i(\delta-\delta_0-\omega t)}=1.
\end{align}
$A=A_0$ gives
\begin{align}\label{A:Omega t}
    0&=A^2-A^2_0=(c^2-c^2_0)\left(1-\frac{\Omega^2}{\omega^2}\right)\to\Omega t_s=n\pi.
\end{align}
Plugging $\Omega t_s=n\pi$ into Eq. \eqref{A:eta bar},
\begin{align}\label{A:omega t}
   [1-(-1)^ne^{-i\omega t}]\left(c_0+i\frac{\Omega}{\omega}s_0\right)=0\to \omega t_s+n\pi=2m\pi.
\end{align}
Eq. \eqref{A:Omega t} and Eq. \eqref{A:omega t} are equivalently written as
\begin{align}\label{A:Omega t omega t}\nonumber
       &\Omega t_s=2n\pi, \omega t_s=2m\pi\\
       \text{or}&\\
        &\Omega t_s=(2n+1)\pi,\omega t_s=(2m+1)\pi,~(n,m\in\mathbb{N}).\nonumber
\end{align}
%%%%%%%%%%%%%%%%%%%%%%%%%%%%%%%%%%%%%%%%%%%%%
\subsection{Beating effect}\label{A:Beating effect}
$Q$ defined in Eq. \eqref{A:Q} is in $1/(\omega-\Omega)$ order. The coefficient in a zeroth order of $\omega$ in $\bar{\eta}$ at $\omega=\Omega$ is zero:
\begin{align}\label{A:eta bar 0th}
    \bar{\eta}_{\omega=\Omega}&=-e^{-i\Omega t}e^{i\Omega t+i\phi}+e^{+i\phi}=0.
\end{align}
we expand $\bar{\eta}$ in Eq. \eqref{A:eta bar} by $\omega$ around $\Omega$:
\begin{align}\label{A:eta Taylor series}
    \bar{\eta}&= \bar{\eta}_{\omega=\Omega}+\left.\frac{\partial\bar{\eta}}{\partial\omega}\right|_{\omega=\Omega}(\omega-\Omega)+\frac{1}{2!}\left.\frac{\partial^2\bar{\eta}}{\partial\omega^2}\right|_{\omega=\Omega}(\omega-\Omega)^2+\cdots.
\end{align}
A first order derivative of $\bar{\eta}$ with respect to $\omega$ is
\begin{align}\label{A:eta bar 1st}\nonumber
    \frac{\partial\bar{\eta}}{\partial\omega}&=-ite^{-i\omega t}\left(c+i\frac{\Omega}{\omega}s\right)+ie^{-i\omega t}\frac{\Omega}{\omega^2}s-i\frac{\Omega}{\omega^2}s_0\\
    &=e^{-i\omega t}\left(-itc+t\frac{\Omega}{\omega}s+i\frac{\Omega}{\omega^2}s\right)-i\frac{\Omega}{\omega^2}s_0.
\end{align}
A second order derivative of $\bar{\eta}$ with respect to $\omega$ is
\begin{align}\label{A:eta bar 2nd}\nonumber
     \frac{\partial^2\bar{\eta}}{\partial\omega^2}&=e^{-i\omega t}\left(-t^2c-it^2\frac{\Omega}{\omega}s+t\frac{\Omega}{\omega^2}s-t\frac{\Omega}{\omega^2}s-2i\frac{\Omega}{\omega^3}s\right)+2i\frac{\Omega}{\omega^3}s_0\\
     &=e^{-i\omega t}\left(-t^2c-it^2\frac{\Omega}{\omega}s-2i\frac{\Omega}{\omega^3}s\right)+2i\frac{\Omega}{\omega^3}s_0.
\end{align}
The coefficient of a first order of $\omega$ in $\bar{\eta}$ at $\omega=\Omega$ is finite:
\begin{align}\label{A:eta bar 1st}\nonumber
    \left.\frac{\partial\bar{\eta}}{\partial\omega}\right|_{\omega=\Omega}
    &=e^{-i\Omega t}\left(-itc+ts+\frac{i}{\Omega}s\right)-\frac{i}{\Omega}s_0\\
    &=-ite^{i\phi}+\frac{i}{\Omega}(se^{-i\Omega t}-s_0).
\end{align}
The coefficient in a second order of $\omega$ in $\bar{\eta}$ at $\omega=\Omega$ is finite:
\begin{align}\label{A:eta bar 2nd}\nonumber
     \left.\frac{\partial^2\bar{\eta}}{\partial\omega^2}\right|_{\omega=\Omega}
     &=e^{-i\Omega t}\left(-t^2c-it^2s-\frac{2i}{\Omega^2}s\right)+\frac{2i}{\Omega^2}s_0\\
     &=-t^2e^{i\phi}-\frac{2i}{\Omega^2}(se^{-i\Omega t}-s_0).
\end{align}
$\eta$ is expressed up to a first order in $(\omega-\Omega)$ as 
\begin{align}\label{A:eta2}
    \eta
    &=\frac{g}{\omega+\Omega}\sqrt{\frac{\omega}{2m}}\left\{-ite^{i\phi}+\frac{i}{\Omega}(se^{-i\Omega t}-s_0)+\frac{1}{2}\left[-t^2e^{i\phi}-\frac{2i}{\Omega^2}(se^{-i\Omega t}-s_0)\right](\omega-\Omega)\right\}+O[(\omega-\Omega)^2].
\end{align}
Hence, $\eta$ is finite at the limit $\omega\to\Omega$ as
\begin{align}\label{A:eta limit}
\lim_{\omega\to\Omega}\eta=ig\sqrt{\frac{1}{8m\Omega^3}}\left[e^{-i\Omega t}\sin(\Omega t+\phi)-\sin\phi-\Omega te^{i\phi}\right].
\end{align}
Although there is no divergence at $\omega=\Omega$, at $\omega\simeq \Omega$ there exists a beating effect. Expressing $\bar{\eta}$ in terms of the Euler's representation,
\begin{align}\label{A:eta bar Euler}\nonumber
\bar{\eta}&=-e^{-i\omega t}\frac{1}{2}\left[e^{i(\Omega t+\phi)}+e^{-i(\Omega t+\phi)}+\frac{\Omega}{\omega}e^{i(\Omega t+\phi)}-\frac{\Omega}{\omega}e^{-i(\Omega t+\phi)}\right]+ \frac{1}{2}\left[e^{i\phi}+e^{-i\phi}+\frac{\Omega}{\omega}e^{i\phi}-\frac{\Omega}{\omega}e^{-i\phi}\right]\\
 &=-e^{-i\omega t}\frac{1}{2}\left[\left(1+\frac{\Omega}{\omega}\right)e^{i(\Omega t+\phi)}+\left(1-\frac{\Omega}{\omega}\right)e^{-i(\Omega t+\phi)}\right]+\frac{1}{2}\left[\left(1+\frac{\Omega}{\omega}\right)e^{i\phi}+\left(1-\frac{\Omega}{\omega}\right)e^{-i\phi}\right],
\end{align}
$|\bar{\eta}|^2$ is written as the linear combination of cosines with different frequencies, $\Omega+\omega$, $2\Omega$ and $\Omega-\omega$: 
\begin{align}\label{A:eta bar absolute cosines}\nonumber
|\bar{\eta}|^2
 &=-\frac{1}{2}\left(1+\frac{\Omega}{\omega}\right)^2\cos(\Omega-\omega)t-\frac{1}{2}\left(1-\frac{\Omega^2}{\omega^2}\right)\cos[(\Omega-\omega)t+2\phi]\\
 &-\frac{1}{2}\left(1-\frac{\Omega}{\omega}\right)^2\cos(\Omega+\omega)t-\frac{1}{2}\left(1-\frac{\Omega^2}{\omega^2}\right)\cos[(\Omega+\omega)t+2\phi]\\
 &+1+\frac{\Omega^2}{\omega^2}+\frac{1}{2}\left(1-\frac{\Omega^2}{\omega^2}\right)[\cos2(\Omega t+\phi)+\cos2\phi].\nonumber
\end{align}
 Here we have the beating relevant part $\cos(\Omega-\omega)t$ and $\cos[(\Omega-\omega)t+2\phi]$. These form a large envelop without making zero within a large timescale and the rest of the cosines with the frequency $(\Omega+\omega)$ take part in micro-motions around $\omega\simeq \Omega$. Eq. \eqref{A:eta bar absolute cosines} can be re-written as 
 \begin{align}\label{A:eta bar absolute cosines 2}\nonumber
  |\bar{\eta}|^2
 &=D\cos(\Delta\omega t-\xi)+ 1+\frac{\Omega^2}{\omega^2}\\
&-\frac{1}{2}\left(1-\frac{\Omega}{\omega}\right)^2\cos(\Omega+\omega)t-\frac{1}{2}\left(1-\frac{\Omega^2}{\omega^2}\right)\cos[(\Omega+\omega)t+2\phi]\\
&+\frac{1}{2}\left(1-\frac{\Omega^2}{\omega^2}\right)[\cos2(\Omega t+\phi)+\cos2\phi],\nonumber
 \end{align}
where $\Delta\omega \equiv|\Omega-\omega|$,
\begin{align}\label{A:eta bar absolute cosines angle}
    \sin\xi\equiv\frac{1}{2D}\left(1-\frac{\Omega^2}{\omega^2}\right)\sin2\phi,~ \cos\xi\equiv-\frac{1}{2D}\left[\left(1+\frac{\Omega}{\omega}\right)^2+\left(1-\frac{\Omega^2}{\omega^2}\right)\cos2\phi\right]
\end{align}
and
\begin{align}\label{A:eta bar absolute cosines beating amp}
    D\equiv\frac{1}{2}\sqrt{2\left(1+\frac{\Omega}{\omega}\right)^2\left(1-\frac{\Omega^2}{\omega^2}\right)\cos2\phi+\left(1+\frac{\Omega}{\omega}\right)^4+\left(1-\frac{\Omega^2}{\omega^2}\right)^2}.
\end{align}
For $\Omega>\omega$, $\phi=\frac{\pi}{2}$ maximizes $|\bar{\eta}|^2$ while $\phi=0$ minimizes $|\bar{\eta}|^2$. For $\Omega<\omega$, $\phi=0$ maximizes $|\bar{\eta}|^2$ while $\phi=\frac{\pi}{2}$ minimizes $|\bar{\eta}|^2$.
\twocolumngrid

\begin{acknowledgments}
The author is sincerely grateful for many valuable and critical comments to Prof. Jaros\l{}aw K. Korbicz.
\end{acknowledgments}
\section*{Availability of data and material}
This work does not include any data and materials.
\section*{Funding}
The author acknowledges the support
from the Polish National Science Center (NCN), Grant No.  2019/35/B/ST2/01896.
\section*{Author information}
\subsection*{Authors and Affiliations}
\noindent
Author: \textbf{Tae-Hun Lee}\\
Affiliation: \textbf{Center for Theoretical Physics, Polish Academy of Sciences, Warsaw, Poland}\\
ORCID:\href{https://orcid.org/0000-0002-9199-3372}{0000-0002-9199-3372}
\subsection*{Contributions}
The author contributed to deriving, analyzing the formulas and writing the manuscript. The author read and approved the final manuscript.
\subsection*{Corresponding authors
}
\noindent
Correspondence to \href{mailto:taehunee@gmail.com}{Tae-Hun Lee}
\section*{Ethics declarations}
\subsection*{Competing interests}
The author declares that there are no competing interests.

\begin{thebibliography}{30}
%%%%%%%%%%%%%%%%%%%%%%%%%%%%%%%%%%%%%%%%%%%%%%%%%%%%%%%%%%%%%1
 %\cite{Zurek:2009ymk},
\bibitem{ZurekNature2009}
W.~H.~Zurek,
%``Quantum Darwinism,''
\href{https://doi.org/10.1038/nphys1202}{Nat. Phys. \textbf{5}, 181-188 (2009)}. %no.3,  
%doi.org/10.1038/nphys1202.
%169 citations counted in INSPIRE as of 11 Jan 2024
%%%%%%%%%%%%%%%%%%%%%%%%%%%%%%%%%%%%%%%%%%%%%%%%%%%%%%
%\cite{{Zurek:2014}}
\bibitem{ZurekPhysToday2014}
W.~H.~Zurek, 
%``Quantum Darwinism, classical reality, and the randomness of quantum jumps,''
\href{https://doi.org/10.1063/PT.3.2550}{Phys. Today \textbf{67}, 44-50 (2014)}. 
%doi.org/10.1063/PT.3.2550.
%%%%%%%%%%%%%%%%%%%%%%%%%%%%%%%%%%%%%%%%%%%%%%%%%%%%
%\cite{Korbicz:2020tkf} 
\bibitem{Korbicz:2014}
 J.~K.~Korbicz, R.~Horodecki, and P.~ Horodecki, 
 %``Objectivity in a Noisy Photonic Environment through Quantum State Information Broadcasting,'' 
 \href{https://doi.org/10.1103/PhysRevLett.112.120402}{Phys. Rev. Lett. \textbf{112}, 120402  (2014)}. %doi:10.1103/PhysRevLett.112.120402
%%%%%%%%%%%%%%%%%%%%%%%%%%%%%%%%%%%%%%%%%%%%%%%%%%%%%
\bibitem{Korbicz:2020tkf}
J.~K.~Korbicz,
%``Roads to objectivity: Quantum Darwinism, Spectrum Broadcast Structures, and Strong quantum Darwinism \textendash{} a review,''
 \href{https://doi.org/10.22331/q-2021-11-08-571}{Quantum \textbf{5}, 571 (2021)}. 
%doi:10.22331/q-2021-11-08-571
%[arXiv:2007.04276 [quant-ph]].
%35 citations counted in INSPIRE as of 11 Jan 2024
%%%%%%%%%%%%%%%%%%%%%%%%%
\bibitem{Riedel:2011}
C.~J. ~Riedel and W. ~H. ~Zurek,
%``Redundant information from thermal illumination: quantum Darwinism in scattered photons''
\href{https://doi.org/10.1088/1367-2630/13/7/073038}{New J. Phys \textbf{13}, 073038 (2011)}. 
%doi:10.1088/1367-2630/13/7/073038
%%%%%%%%%%%%%%
\bibitem{Riedel:2012}
C.~J. ~Riedel, W. ~H. ~Zurek and M. ~Zwolak
%``The rise and fall of redundancy in decoherence and quantum Darwinism''
\href{https://doi.org/10.1088/1367-2630/14/8/083010}{New J. Phys \textbf{14}, 083010 (2012)}.
%doi:10.1088/1367-2630/14/8/083010
%%%%%%%%%%%%%%

%%%%%%%%%%%%%%
\bibitem{Zwolak:2013}
M. ~Zwolak and W. ~H. ~Zurek
%``Complementarity of quantum discord and classically accessible information''
\href{https://doi.org/10.1038/srep01729}{Sci. Rep. \textbf{3}, 1729 (2013)}.
%doi:10.1038/srep01729
%%%%%%%%%%%%%%
\bibitem{Zwolak:2016}
M. ~Zwolak, C.~J. ~Riedel and W. ~H. ~Zurek
%``Amplification, Decoherence and the Acquisition of Information by Spin Environments''
\href{https://doi.org/10.1038/srep25277}{Sci. Rep. \textbf{6}, 25277 (2016)}.
%doi:10.1038/srep25277
%%%%%%%%%%%%%%%
%\cite{Le:2019mnn}
\bibitem{Le:2019mnn}
T.~P.~Le and A.~Olaya-Castro,
%``Strong Quantum Darwinism and Strong Independence is equivalent to Spectrum Broadcast Structure,''
\href{https://doi.org/10.1103/PhysRevLett.122.010403}{Phys. Rev. Lett. \textbf{122} 010403 (2019)}
%73 citations counted in INSPIRE as of 10 Mar 2026
%%%%%%%%%%%%%%
\bibitem{Mironowicz:2017}
P. ~Mironowicz, J.~K. ~Korbicz, and P. ~Horodecki
%``Monitoring of the Process of System Information Broadcasting in Time''
\href{https://doi.org/10.1103/PhysRevLett.118.150501}{
Phys. Rev. Lett. \textbf{118}, 150501 (2017)}.
%%%%%%%%%%%%%%%%%%%%%%%%%%%%%%%%%%%%%%%%%%%%%%%%%%%%%%%%%%%
\bibitem{Mironowicz:2018}
P. ~Mironowicz, P. ~Nale\.{z}yty, P. ~Horodecki, and J. ~K. ~Korbicz
%``System information propagation for composite structures''
\href{https://doi.org/10.1103/PhysRevA.98.022124}{
Phys. Rev. A \textbf{98}, 02212 (2018)}.
%%%%%%%%%%%%%%%%%%%%%%%%%%%%%
\bibitem{Lampo:2017}
A.~Lampo, J. ~Tuziemski, M.~ Lewenstein, and J. ~K. ~Korbicz
%``Objectivity in the non-Markovian spin-boson model''
\href{https://doi.org/10.1103/PhysRevA.96.012120}{
Phys. Rev. A \textbf{96}, 012120 (2017)}.
%%%%%%%%%%%%%%%%%%%%%%%%%%%%%%%%%%
%\cite{Lee:2024iso}
\bibitem{Lee:2024iso}
T.~H.~Lee and J.~K.~Korbicz,
%``Encoding position by spins: Objectivity in the boson-spin model,''
\href{https://doi.org/10.1103/PhysRevA.109.052204}{Phys. Rev. A \textbf{109}, 052204 (2024)}.
%[arXiv:2401.07690 [quant-ph]].
%2 citations counted in INSPIRE as of 14 Jun 2025
%%%%%%%%%%%%%%%%%%%%%%%%%
%\cite{Lee:2024idx}
\bibitem{Lee:2024idx}
T.~H.~Lee and J.~K.~Korbicz,
%``Holevo bound and objectivity in the boson-spin model,''
\href{https://doi.org/10.1103/PhysRevA.110.062202}{
Phys. Rev. A \textbf{110}, 062202 (2024)}.
%[arXiv:2409.01186 [quant-ph]].
%0 citations counted in INSPIRE as of 14 Jun 2025
%%%%%%%%%%%%%%%%%%%%%%%%%%%%%%%%%%%%%%%%%%%%%%%%%%
\bibitem{Tuziemski:2015}
J.~Tuziemski and J. ~K. ~Korbicz, 
%``Dynamical objectivity in quantum Brownian motion'', 
\href{https://doi.org/10.1209/0295-5075/112/40008}{Europhys. Lett. \textbf{112}, 40008 (2015)}.
%%%%%%%%%%%%%%%%%%%%%%%%%%%%%%%%%%%%%%%%%%%%%%%%%%%
%\cite{Lee:2023ozm}
\bibitem{Lee:2023ozm}
T.~-H.~Lee and J.~K.~Korbicz,
%``Complementarity between decoherence and information retrieval from the environment,''
\href{https://doi.org/10.1103/PhysRevA.109.032221}{Phys. Rev. A \textbf{109}, 032221 (2024)}.
%3 citations counted in INSPIRE as of 02 Apr 2025
%%%%%%%%%%%%%%%%%%%%%%%%%%%%%%%%%%%%%%%%%%%%%%%%%%%%%%
\bibitem{Schlosshauer2007}
M.~Schlosshauer, 
\href{https://doi.org/10.1007/978-3-540-35775-9}{{\textit{Decoherence and the Quantum-to-Classical Transition}}} (Springer, Berlin, 2007).
%%%%%%%%%%%%%%%%%%%%%%%%%%%%%%%%%%%%%%%%%%%%%%%%%%%%%%
\bibitem{Born:1927}
M. ~Born and J. ~R. ~Oppenheimer,
%``Zur Quantentheorie der Molekeln,''
\href{ https://doi.org/10.1002/andp.19273892002}{Ann. Phys. \textbf{389}, 457 (1927)}.
%%%%%%%%%%%%%%%%%%%%%%%%%%%%%%%%%%%%%%%%%%%%%%%%%%%%%
\bibitem{Ullersma:1966}
P. ~Ullersma,
%``An exactly solvable model for Brownian motion: IV. Susceptibility and Nyquist's theorem,''
\href{https://doi.org/10.1016/0031-8914(66)90105-4}{Physica \textbf{32}, 90-96 (1966)}.
%%%%%%%%%%%%%%%%%%%%%%%%%%%%%%%%%%%%%%%%%%%%%%%%%
\bibitem{Joos:2003}
E.~Joos, H. ~D.~Zeh , C.~Kiefer, D. ~Giulini, J. ~Kupsch, I.~-O. ~Stamatescu,
\href{https://doi.org/10.1007/978-3-662-05328-7}{\textit{Decoherence and the Appearance of a Classical World in Quantum Theory} (Springer, Berlin, 2003)}.
%%%%%%%%%%%%%%%%%%%%%%%%%%%%%%%%%%%%%%%%%%%%%%%%%%%%%
\bibitem{Petruccione:2010}
H.~-P. ~Breuer and F.~Petruccione,
\href{https://10.1093/acprof:oso/9780199213900.001.0001}{\textit{The Theory of Open Quantum Systems} (Oxford University Press, 2010)}.
%%%%%%%%%%%%%%%%%%%%%%%%%%%%%%%%%%%%%%%%%%%%%%%%%%%
\bibitem{Paz:2009}
J.~P.~Paz and A.~J.~Roncaglia
%``Redundancy of classical and quantum correlations during decoherence''
\href{https://doi.org/10.1103/PhysRevA.80.042111}{Phys. Rev. A \textbf{80}, 042111 (2009)}.
%doi:https://doi.org/10.1209/0295-5075/112/40008
%%%%%%%%%%%%%%%%%%%%%%%%%%%%%%%%%%%%%%%%%%%%%%%%%%%%%%%%%
\bibitem{Ingold:2002}
G.~-L. ~Ingold,
\href{https://doi.org/10.1007/3-540-45855-7_1}{\textit{Path Integrals and Their Application to Dissipative Quantum Systems In: A.~Buchleitner, K.~Hornberger (eds) Coherent Evolution in Noisy Environments}}, Lecture Notes in Physics, vol 611 (Springer, Berlin, Heidelberg, 2002), pp 1-53.
%%%%%%%%%%%%%%%%%%%%%%%%%%%%%%%%%%%%%%%%%%%%%%%%%%%%%%
\bibitem{Floquet:1883}
G. ~Floquet,  
%''Sur les \'equations diff\'erentielles lin\'eaires  \`a coefficients p\'eriodiques'', 
\href{https://www.numdam.org/articles/10.24033/asens.220/}{
Ann. Sci. \'Ec. Norm. Sup\'er. (2), \textbf{12}, 47 (1883)}.
\bibitem{Husimi:1953}
K.~Husimi, 
%``Miscellanea in Elementary Quantum Mechanics, II'', 
\href{https://doi.org/10.1143/ptp/9.4.381}{Prog. Theor. Phys \textbf{9}, 381-402 (1953)}.
%%%%%%%%%%%%%%%%%%%%%%%%%%%%%%%%%%%%%%%%%%%%%%%%%%%
\bibitem{Hanggi:2016}
P. ~H\"anggi,
\href{https://www.physik.uni-augsburg.de/theo1/hanggi/Chapter_5.pdf}{Driven Quantum Systems (2016).}
%%%%%%%%%%%%%%%%%%%%%%%%%%%%%%%%%%%%%%%%%%%%%%%%%%%%%%%
\bibitem{Bukov:2017}
M. ~G. ~ Bukov,
%``Floquet engineering in periodically driven closed quantum systems: from dynamical localisation to ultracold topological matter,''  
\href{https://hdl.handle.net/2144/43881}{[PhD's thesis, Boston University] (2017)}.
%%%%%%%%%%%%%%%%%%%%%%%%%%%%%%%%%%%%%%%%%%%%%%%%
\bibitem{Helstrom:1969}
C. ~W. ~ Helstrom,
%``Quantum detection and estimation theory,''  
\href{https://doi.org/10.1007/BF01007479}{J. Stat. Phys. \textbf{1}, 231-252 (1969)}.
%%%%%%%%%%%%%%%%%%%%%%%%%%%%%%%%%%%%%%%%%%%%%%%%%%%%%%%
%\cite{Sudarshan:1963ts}
\bibitem{Sudarshan:1963ts}
E.~C.~G.~Sudarshan,
%``Equivalence of semiclassical and quantum mechanical descriptions of statistical light beams,''
\href{https://link.aps.org/doi/10.1103/PhysRevLett.10.277}
{{Phys. Rev. Lett. \textbf{10}, 277-279 (1963)}}.
%805 citations counted in INSPIRE as of 28 Mar 2025
%%%%%%%%%%%%%%%%%%%%%%%%%%%%%%%%%%%%%%%%%%%%%%%%%%%%%%%%%%%%%1
\bibitem{Mehta:1967}
C. ~L.~Mehta,
%``Diagonal Coherent-State Representation of Quantum Operators,''
\href{https://link.aps.org/doi/10.1103/PhysRevLett.18.752}
{{Phys. Rev. Lett. \textbf{18}, 752--754 (1967)}}.
%%%%%%%%%%%%%%%%%%%%%%%%%%%%%%%%%%%%%%%%%%%%%%%%%%%%%%%
\end{thebibliography}
\end{document}